\documentclass[sigconf, authorversion]{acmart}

\usepackage{dblfloatfix}
\usepackage{enumitem}
\usepackage{colortbl}
\usepackage{multirow}
\usepackage{makecell}
\usepackage{booktabs}

\newcommand{\accessdate}{ (Accessed May 6, 2025)}

\newcommand{\LowLatency}{\textit{Low (1.5s)}}
\newcommand{\MediumLatency}{\textit{Medium (4.0s)}}
\newcommand{\HighLatency}{\textit{High (6.5s)}}

\newcommand{\PerceivedLatency}{Response Time}
\newcommand{\PerceivedLatencyLowercase}{response time}

\AtBeginDocument{%
  \providecommand\BibTeX{{%
    \normalfont B\kern-0.5em{\scshape i\kern-0.25em b}\kern-0.8em\TeX}}}

\setcopyright{acmlicensed}\acmConference[CUI '25]{Proceedings of the 7th ACM Conference on Conversational User Interfaces}{July 8--10, 2025}{Waterloo, ON, Canada}
\copyrightyear{2025}
\acmYear{2025}
\acmBooktitle{Proceedings of the 7th ACM Conference on Conversational User Interfaces (CUI '25), July 8--10, 2025, Waterloo, ON, Canada}
\acmDOI{10.1145/3719160.3736636}
\acmISBN{979-8-4007-1527-3/2025/07}

\begin{document}

\title[Conversational Agents' Response Latency Mitigation]{Mitigating Response Delays in Free-Form Conversations with LLM-powered Intelligent Virtual Agents}

\author{Mykola Maslych}
\orcid{0000-0001-7037-3513}
\affiliation{
  \institution{University of Central Florida}
  \city{Orlando}
  \state{Florida}
  \country{USA}
}
\email{mykola.maslych@ucf.edu}

\author{Mohammadreza Katebi}
\orcid{0009-0007-6752-5628}
\affiliation{
  \institution{NeuroVulpis}
  \city{Orlando}
  \state{Florida}
  \country{USA}
}
\email{m.r.katebii@gmail.com}

\author{Christopher Lee}
\orcid{0009-0000-8059-1741}
\affiliation{
  \institution{Virginia Tech}
  \city{Blacksburg}
  \state{Virginia}
  \country{USA}
}
\email{chrislee24@vt.edu}

\author{Yahya Hmaiti}
\orcid{0000-0003-1052-1152}
\affiliation{
  \institution{University of Central Florida}
  \city{Orlando}
  \state{Florida}
  \country{USA}
}
\email{Yohan.Hmaiti@ucf.edu}

\author{Amirpouya Ghasemaghaei}
\orcid{0009-0002-8463-1207}
\affiliation{
  \institution{University of Central Florida}
  \city{Orlando}
  \state{Florida}
  \country{USA}
}
\email{aghaei.ap@gmail.com}

\author{Christian Pumarada}
\orcid{0009-0002-6756-7442}
\affiliation{
  \institution{University of Central Florida}
  \city{Orlando}
  \state{Florida}
  \country{USA}
}
\email{cpuma@ucf.edu}

\author{Janneese Palmer}
\orcid{0009-0005-0328-0532}
\affiliation{
  \institution{University of Central Florida}
  \city{Orlando}
  \state{Florida}
  \country{USA}
}
\email{ja448612@ucf.edu}

\author{Esteban Segarra Martinez}
\orcid{0000-0002-9412-8464}
\affiliation{
  \institution{University of Central Florida}
  \city{Orlando}
  \state{Florida}
  \country{USA}
}
\email{esteban.segarra@ucf.edu}

\author{Marco Emporio}
\orcid{0000-0002-7601-2935}
\affiliation{
  \institution{University of Verona}
  \city{Verona}
  \country{Italy}
}
\email{marco.emporio@univr.it}

\author{Warren Snipes}
\orcid{0009-0004-8995-4034}
\affiliation{
  \institution{University of Central Florida}
  \city{Orlando}
  \state{Florida}
  \country{USA}
}
\email{warren.snipes@ucf.edu}

\author{Ryan P. McMahan}
\orcid{0000-0001-9357-9696}
\affiliation{
  \institution{Virginia Tech}
  \city{Blacksburg}
  \state{Virginia}
  \country{USA}
}
\email{rpm@vt.edu}

\author{Joseph J. LaViola Jr.}
\orcid{0000-0003-1186-4130}
\affiliation{
  \institution{University of Central Florida}
  \city{Orlando}
  \state{Florida}
  \country{USA}
}
\email{jlaviola@ucf.edu}

\renewcommand{\shortauthors}{Maslych et al.}

\begin{abstract}
We investigated the challenges of mitigating response delays in free-form conversations with virtual agents powered by Large Language Models (LLMs) within Virtual Reality (VR). For this, we used conversational fillers, such as gestures and verbal cues, to bridge delays between user input and system responses and evaluate their effectiveness across various latency levels and interaction scenarios. We found that latency above 4 seconds degrades quality of experience, while natural conversational fillers improve perceived response time, especially in high-delay conditions. Our findings provide insights for practitioners and researchers to optimize user engagement whenever conversational systems' responses are delayed by network limitations or slow hardware. We also contribute an open-source pipeline that streamlines deploying conversational agents in virtual environments.
\end{abstract}



\begin{teaserfigure}
    \centering
    \includegraphics[width=.75\textwidth]{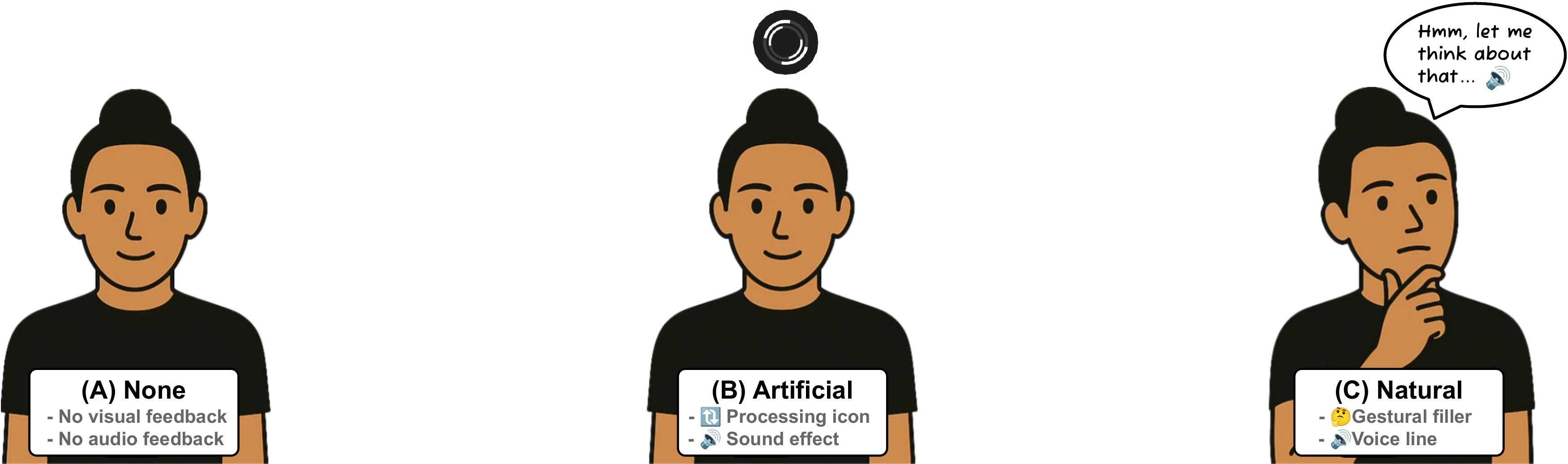}
    \caption{In this paper, we investigated response delays with intelligent virtual agents and the effects of conversational fillers: (A)~\textit{None}:~The user receives no feedback while waiting on the agent; (B)~\textit{Artificial}:~The user receives a processing icon and sound effect; and (C)~\textit{Natural}:~The user receives feedback in the form of social interaction cues (gesture and voice).}
    \Description{Teaser figure gives a high-level overview of what the conversational filler conditions present in the study. The subfigure (A) to the left shows an agent in the none filler condition, where the user does not receive any feedback while waiting on the agent to respond. The middle subfigure (B) shows an agent in the artificial filler condition, such that the user receives a processing icon and sound effect as feedback. The right subfigure (C) shows an agent in the natural filler condition, where the user receives feedback in the form of social interaction cue.}
    \label{fig:teaser}
\end{teaserfigure}


\maketitle

\section{Introduction}

Advancements in artificial intelligence (AI), particularly in large language models (LLMs), automatic speech recognition (ASR) and text-to-speech (TTS), catalyzed the integration of intelligent virtual agents (IVAs) into mainstream systems~\cite{shoa_sushi_2023, wang_virtuwander_2024, wan_building_2024}. These advancements make IVAs more accessible and capable than ever before, evolving beyond simple scripted responses. They offer personalized interactions that adapt in real-time, recall past conversations, and provide a truly realistic and engaging user experience (UX). The usefulness of these intelligent agents extends beyond entertainment
~\cite{wan_building_2024, casas_moodflow_2024, park_generative_2023} to encompass 
medical~\cite{ali_virtual_2020, saeed_developing_2024, llanes_jurado_developing_2024, chheang_towards_2024, steenstra_virtual_2024}, 
educational~\cite{petersen_pedagogical_2021, wang_virtuwander_2024, hasan_sapien_2023, divekar_foreign_2022, topsakal_framework_2022}, 
personal~\cite{zhu_free_form_2023, yamazaki_open-domain_2023, aneja_understanding_2021} and 
social~\cite{shoa_sushi_2023, qin_charactermeet_2024} applications.

More recent applications of IVAs emphasize the support of free-from conversation. To achieve this, most modern architectures for IVAs rely on TTS to vocalize LLM-generated responses to speech transcribed with ASR~\cite{llanes_jurado_developing_2024, yamazaki_open-domain_2023, wang_virtuwander_2024, shoa_sushi_2023, wan_building_2024, bayat_exploring_2024, casas_moodflow_2024, zhu_free_form_2023, hasan_sapien_2023, maslych2024takeaways, garcia_speaking_2024, srinidhi2024xair}. These components are computationally demanding, and when running locally, they compete for system resources with graphics and physics simulations~\cite{muller_brockhausen_chatter_2023}, leading to system response time (SRT) that exceeds the natural duration of silence between interlocutors' turns. To alleviate this, most prior architectures offload processing to cloud-based models. However, cloud solutions are inherently unpredictable, prone to network congestion, connection instability, and system downtime, all of which exacerbate latency. Whether due to hardware limitations, cloud infrastructure, or local system contention, conversational systems' response latency remains a significant challenge, degrading Quality of Experience (QoE)~\cite{shaikh_quality_2010, laghari_application_2019, mahmud_quality_2019}.

Long response times from virtual humans, robots, and voice assistants cause impatience~\cite{niculescu2010socializing, yang2015effect}, frustration~\cite{kum_can_2022, yang2015effect, mcwilliams_secondary_2015}, and general dissatisfaction~\cite{boukaram2021mitigating, funk2020usable, hone2006empathic}. However, findings on response latency and its mitigation in conversations with virtual agents relied on scripted questions and answers~\cite{boukaram2021mitigating, kum_can_2022, jeong2019exploring}, used WoZ protocol~\cite{wigdor2016improve, funakoshi_smoothing_2008, shiwa_how_2009}, or rated pre-recorded conversations~\cite{ohshima_conversational_2015, lopez2019testing}, limiting their generalization to real-time interactions. Addressing this, our work is the first to examine the effects of response latency with \textit{truly interactive} conversational agents powered by LLMs to support free-form conversations, and we aim to answer the following research questions: 
\textbf{(1)} \textit{Do prior findings on perception of latency remain consistent when using free-form conversational IVAs?} 
\textbf{(2)} \textit{Do natural conversational fillers mitigate the negative effects of latency when interacting with free-form IVAs?} 
\textbf{(3)} \textit{Do artificial wait indicators mitigate the negative effects of latency when interacting with free-form IVAs?}

We investigated the effects of response latency and conversational fillers on perceived response time and broader dimensions of user experience in free-form task-guided voice conversations with LLM-based IVAs in Virtual Reality (VR). Participants experienced three virtual worlds, conversing with a total of 9 virtual agents under three filler types: \textit{None}, \textit{Artificial}, and \textit{Natural} (see~\autoref{fig:teaser}), along with three response latency levels: \LowLatency, \MediumLatency, and \HighLatency{} (see~\autoref{sec:delay_levels}). We used an LLM and TTS to generate responses to users' speech transcribed with ASR in real-time. Latency was simulated by delaying the agents' response onset to study the effect of different conversational fillers and response latencies on users' perception of the agents.

We found that delay significantly worsened participants' perceived response time and broader perception metrics, and was less bearable beyond 4 seconds. \textit{Natural} conversational fillers mitigated some of these effects, however \textit{Artificial} wait indicators did not significantly affect user experience. To streamline voice-based conversational agent studies, we provide an open-source library\footnote{\href{https://github.com/ISUE/iva-cui}{github.com/ISUE/iva-cui}}, which includes a system that uses natural language processing, capable of real-time analysis of conversation history to identify and manage task transitions seamlessly. Overall, our contributions inform the design of IVAs and embodied conversational agents (ECAs) applicable to inherently virtual, as well as physical agents, digital twins of which can be rendered in VR.

\section{Related Work}
\label{sec:related_work}

Latency in conversations (response delay) refers to the delay between one speaker's utterance and the other's response. In human communication, brief delays and silences are natural~\cite{clark1996using, jaffe1970rhythms}, yet research suggests that response times beyond two seconds begin to feel unnatural, and silences exceeding four seconds can disrupt conversational flow, signaling a breakdown in communication~\cite{miller_response_1968}. 
Studies on turn-taking in vocal conversations report sub-second latency on average~\cite{heldner_pauses_2010, kanda_humanoid_2007, stivers_universals_2009}, and show that faster response times signal greater social connection~\cite{templeton_fast_2022}. They also show that in natural dialogues, some responses begin even before the previous speaker has finished~\cite{heldner_pauses_2010, stivers_universals_2009, levinson_timing_2015}. In conversational user interfaces, fast, seamless interactions improve engagement, while excessive delays can cause frustration, disrupt the conversational flow, and reduce trust in the system. This underscores the importance of minimizing response latency in human-computer interactions.

When designing virtual agents that rely on ASR, LLMs, and TTS systems, latency poses an even greater challenge due to computational constraints. While techniques like incremental response generation~\cite{tsai2019faster, skantze2013towards} exist, they are not always viable --- a useful response often depends on the information contained at the end of the user's turn, which has not been fully processed before the agent starts responding. Considering delays caused by external factors unrelated to the process of response generation, we focus on latency mitigation through UI-based strategies, drawing from research in domains where delays are an inherent constraint: web and mobile user interfaces, text-based chat interfaces, and embodied human-agent (HAI) and human-robot interactions (HRI). In the following subsections, we outline the strategies used to reduce the perceived delay across various interface types.

\subsection{Latency Mitigation in 2D Interfaces}
\label{sec:related_classical}

Overestimation of wait times by users is a persistent UI design concern~\cite{hornik1984subjective}, so traditional research on delays in human-computer interfaces recommends always displaying a visual progress indicator to show the system is processing a task~\cite{myers1985importance}.

In web browser applications, user satisfaction significantly decreases when response times exceed 12 seconds, often leading users to abandon the application~\cite{hoxmeier2000system}. To address this, some research focused on progress bars behaviors, which significantly impact users' perceived wait times on websites~\cite{harrison2007rethinking, branaghan2009feedback}, although results regarding the optimal progress indicator speed are mixed. Some work shows that indicators that start fast and end slow induce encouragement, reducing abandonment rates~\cite{conrad2010impact,villar2013meta,kim2017effect} and increasing satisfaction rate~\cite{li2019effect}. Faster initial countdown intervals~\cite{komatsu2024waiting} and animated ribbing moving backward while decelerating~\cite{harrison2010faster} can reduce perceived wait duration. Conversely, other work found that users are more tolerant of initial negative progress with an observed preference for linear progress bars, which lead to shorter perceived wait times~\cite{harrison2007rethinking, branaghan2009feedback, amer2016information}. Furthermore, interactive animations substantially reduce perceived wait times by distracting user attention, making wait times feel shorter, and increasing user satisfaction in contrast with passive animations and standard progress bars~\cite{hohenstein2016shorter}. 

In mobile applications, interactive and color-changing loading screens improve user satisfaction by shrinking perceived wait times compared to passive ones~\cite{cheng2024effects}. A study comparing bar and cartoon-bar indicators at constant, accelerating, and decelerating speeds found that decelerating progress bars reduced perceived wait time, and cartoon-bars increased user acceptance and satisfaction~\cite{li2019effects}.

\subsection{Latency in Text-based Chat Interfaces}

Much research on human-to-human text chats and instant messaging (IM) has explored response latency. Users prefer seeing a typing indication (speech bubble or live typing) over nothing while waiting for a reply~\cite{iftikhar2023together}. However, under time pressure, users rated their partners as less involved, reporting frustration even in the presence of a typing indicator~\cite{hwang_when_2019}. Interestingly, third-party observers were more forgiving of delayed responses in support chats, as long as the conversation felt contingent~\cite{lew_interactivity_2018}. While typos are common in human messages, chatbots were rated as less human when their replies contained typos in a WoZ study~\cite{westerman_i_2019}.

In text chatbot interfaces, users prefer short delay over long delay or zero delay, because they perceive it as more realistic~\cite{appel2012does,gnewuch2018faster,shechtman2003media,holtgraves_perceiving_2007}, and presence of a typing indicator during response generation increased the feeling of social presence for novice users~\cite{gnewuch2018chatbot}. Inspired by chatbot interfaces familiar to users, LLM-powered text chat interfaces adopted similar approaches to mitigate response delays, accompanied by the live appearance of response text as it is being generated. Systems like
ChatGPT\footnote{\href{https://chatgpt.com/}{chatgpt.com}\accessdate}, 
Claude\footnote{\href{https://www.anthropic.com/claude}{anthropic.com/claude}\accessdate}, 
Gemini\footnote{\href{https://gemini.google.com/app}{gemini.google.com}\accessdate}, 
MetaAI\footnote{\href{https://www.meta.ai/}{meta.ai}\accessdate}, and 
Perplexity AI\footnote{\href{https://perplexity.ai}{perplexity.ai}\accessdate}
employ distinct loading feedback mechanisms to enhance user experience: 
ChatGPT uses a pulsating black dot;
Claude employs a pulsing star-like object;
Gemini utilizes progress indicators including a rotating star and sequential progress bars;
Perplexity AI uses a rotating ellipsis resembling pages of a book;
MetaAI integrates a spinning wheel.
These techniques inform the user about the ongoing process of response generation, potentially reducing the perceived wait time.

\subsection{Latency Mitigation for Embodied Agents}
\label{sec:related_embodied}
An embodied agent interacts with its physical/virtual environment through physical/virtual body and, unlike text-based chat interfaces, must manage both verbal and visual cues to maintain engagement despite response delays. Excessively long latencies negatively impact UX and users' perception of virtual agents and robots, often leading to user discomfort~\cite{funk2020usable, baylor2006interface, hone2006empathic, boukaram2021mitigating} and assumptions that an error has occurred~\cite{funk2020usable}. Wait times of 4.5 to 5 seconds can also be thought of as negative responses~\cite{porcheron_voice_2018}. During task-solving with robots, delays lead to increased frustration and anger, and decreased satisfaction and future use intention~\cite{yang2015effect}. Additionally, quicker rather than delayed responses from a robot receptionist lead to increased tolerance and reported interaction quality in users~\cite{niculescu2010socializing}.

Research on mitigating response delays has identified conversational fillers~\cite{gambino2017beyond,lopez2019testing,jeong2019exploring, shiwa_how_2009} common in everyday speech as a promising direction. These fillers (e.g. `uh...' and `uhm...') serve several para-linguistic functions~\cite{bevnuvs2014prosody}, including turn-taking management~\cite{svartvik1990london}, social approval~\cite{christenfeld1995does}, and co-creation of pragmatic and discourse~\cite{swerts1998filled}, making them essential for smooth and successful natural conversations~\cite{bortfeld2001disfluency, taboada2006spontaneous}. While simple fillers can reduce a robot's perceived intelligence and likability~\cite{jeong2019exploring},  more complex ones mitigate delay effects without harming perceptions of intelligence~\cite{wigdor2016improve} or virtual agent's competence~\cite{kum_can_2022}. These complex fillers include \textit{pensive fillers} (e.g., `let me think' paired with gestures like chin-scratching) and \textit{acknowledgment fillers} (e.g., `aha' with rapid head-nodding). Systems using such fillers --- whether generic ("uh") or context-aware ("I’ve found a flight for you") -- were rated as more appropriate than silent systems, even with equal delay~\cite{lopez2019testing}.

While the literature is clear on the negative effects of latency on UX, studies in perception of conversational agents relied on predetermined sets of questions and responses~\cite{boukaram2021mitigating, kum_can_2022, jeong2019exploring}, followed WoZ protocol~\cite{wigdor2016improve, funakoshi_smoothing_2008, shiwa_how_2009}, or rated pre-recorded conversations~\cite{ohshima_conversational_2015, lopez2019testing}. This limits their generalization to cases where virtual agents are truly interactive, afforded by recent advancements in the speed and quality of speech and text processing models. Addressing this, we explored how latency affects the perceived virtual agent responsiveness and whether its negative effects can be mitigated by conversation fillers through a study where agents responded to any free-form queries in real-time across multiple scenarios.

\section{Methodology}

This study involved immersive VR scenarios where participants interacted with virtual agents using their speech, under varied response delays and multiple delay mitigation strategies. We selected VR as a medium for our study to make it easily reproducible~\cite{pan_why_2018} and to reduce external factors and distractions~\cite{kuvar_novel_2024}, focusing on variables of interest~\cite{creem_regehr_perceiving_2022}. We expect research on embodied conversational agents to continue expanding, given the rise of VR social applications where users interact with embodied virtual avatars~\cite{maslych2024selectionsdatabase}, and indications that immersive VR induces greater social influence from virtual characters compared to standard desktop applications~\cite{bailey_virtual_2019, kyrlitsias_social_2022}. The remainder of this section outlines key components of the experiment design, conditions, system implementation, apparatus, participants, and the collected data.

\subsection{Experiment Design}
\label{sec:experiment_design}

The experiment included three delay levels: \LowLatency, \MediumLatency, \HighLatency, and three latency mitigation types: \textit{None}, \textit{Artificial Wait Indicator}, \textit{Natural Conversational Filler}, leading to a total of 9 conditions. This matched the number of IVAs present in our study. The order of IVAs was the same across all participants, and the order of conditions was counter-balanced using Balanced Latin Square (18 orders). This way, the conditions applied to IVAs varied across participants.

\subsubsection{Delay Levels}
\label{sec:delay_levels}

Our design included three delay levels: (1) \textit{Fast} at 1.5 seconds, (2) \textit{Medium} at 4.0 seconds, (3) \textit{Slow} at 6.5 seconds. Under \LowLatency{} latency, responses were played as soon as they were generated (SRT average). The longer delays increased in 2.5-second steps: \MediumLatency{} matched the comfortable silence threshold reported in prior work, while \HighLatency{} exceeded it (see~\autoref{sec:related_work}).

\subsubsection{Filler Types}
\label{sec:filler_types}

For latency mitigation, we used three filler types: (1) \textit{None} with no filler; (2) \textit{Artificial Wait Indicator (WI)} with a loading icon and sound; and (3) \textit{Natural Conversational Filler (CF)} with a thinking gesture and a voice line (see~\autoref{fig:teaser}). Both \textit{Artificial} and \textit{Natural} fillers combined visual and auditory cues. We treated these as unified conditions, as prior work showed that multimodal fillers work outperform unimodal ones (see ~\autoref{sec:related_embodied}). Fillers appeared during the wait period (varying by delay level) between the end of participant speech and the agent's response.

In the \textit{None} condition (baseline), agents provided no visual or auditory cues during the delay. They remained in the `attentive idle' animation, facing the participant, and began speaking once the delay elapsed.

In the \textit{Artificial} condition, a rotating visual indicator (concentric quarter-circles) appeared above the agent's head, accompanied by a continuous processing sound (similar to the ChatGPT mobile app). The agents stayed in the `attentive idle' state throughout. Both the icon and sound ended when the delay expired.

In the \textit{Natural} condition, each agent randomly selected from three `thinking' gestures and six filler voice lines at runtime. To simulate deliberation, agents turned their head, touched their chin or the back of their head, and said a conversational filler. They held this pose with subtle breathing motion until the delay ended, then returned to `attentive idle' before responding. Fillers were inspired by prior work~\cite{kum_can_2022, wigdor2016improve} and included: 
"Hmm, let's see...",
"Okay, hmm...",
"Uhmmm...",
"Ah...",
"Hmm, one moment...",
"Hmmm...".

\subsubsection{Hypotheses}
\label{sec:hypotheses}

We derived six hypotheses from prior work: (\textbf{H1a}, \textbf{H1b}) on the negative effects of high latency on user experience~(\autoref{sec:related_work}); (\textbf{H2a}, \textbf{H2b}) on the benefits of conversational fillers in mitigating perceived delay~(\autoref{sec:related_embodied}); and (\textbf{H3a}, \textbf{H3b}) on the role of loading indicators in improving perceived latency in classical UIs~(\autoref{sec:related_classical}). These hypotheses guided our investigation into how latency and conversational fillers affect the perception of embodied IVAs responding to free-form queries in immersive VR:

\begin{enumerate}[label={}, align=left]
\small
    \item[\textbf{H1a:}] Latency degrades perceived \PerceivedLatencyLowercase{} of conversational agents.
    
    \item[\textbf{H1b:}] Latency degrades broader perception dimensions of conversational agents.

    \item[\textbf{H2a:}] Under latency, \textit{Natural} conversational fillers improve perceived \PerceivedLatencyLowercase{} of conversational agents.

    \item[\textbf{H2b:}] Under latency, \textit{Natural} conversational fillers improve broader perception dimensions of conversational agents.
    
    \item[\textbf{H3a:}] Under latency, \textit{Artificial} wait indicators improve perceived \PerceivedLatencyLowercase{} of conversational agents.
    
    \item[\textbf{H3b:}] Under latency, \textit{Artificial} wait indicators improve broader perception dimensions of conversational agents.
\end{enumerate}

\subsection{Study Questionnaires}

To test our hypotheses and gather additional insights into user perception, we created two custom questionnaires informed by prior literature, as no standardized survey exists for interface latency and its mitigation. The first was administered after each condition, and the second after completing all experimental conditions.

\subsubsection{Post-condition Questions}

After each agent interaction, participants completed an in-VR survey by selecting from five labeled response options: \textit{Agree}, \textit{Somewhat Agree}, \textit{Neutral}, \textit{Somewhat Disagree}, and \textit{Disagree}. Q1 assessed perceived response latency and included the word "meaningfully" to avoid bias toward \textit{Natural} fillers that included speech. Q2–Q6 measured other perception dimensions, drawn from prior literature~\cite{kum_can_2022, jeong2019exploring, lopez2019testing, carpinella_robotic_2017} and our study objectives. Q4 and Q5 were adapted from the Robotic Social Attributes Scale (RoSAS)~\cite{carpinella_robotic_2017}, corresponding to discomfort and competence dimensions.

\begin{enumerate}[label=\textbf{(Q\arabic*)}, align=left]
\small
    \item \textbf{\PerceivedLatency}: From the moment I stopped talking, the agent was quick to start responding meaningfully.
    \item \textbf{Engagement}: I felt absorbed during my interaction with this virtual agent.
    \item \textbf{Good Impression}: The virtual agent left a good impression on me.
    \item \textbf{Discomfort}: I felt awkward, scared, and strange when talking to this agent.
    \item \textbf{Competence}: This agent was reliable, competent, and interactive.
    \item \textbf{Willingness to Interact Again}: I would be willing to interact and spend time with this virtual agent again.
\end{enumerate}

\subsubsection{Post-study Questions}

After completing the VR experience, participants filled out a post-study survey (\autoref{tab:post_study}). The first six questions assessed their overall impressions of the agents and whether they noticed the study conditions (i.e., filler types). Participants then watched a 40-second video showing a single agent performing each of the three fillers at \MediumLatency{} to clarify differences in case they had gone unnoticed. All participants viewed the same video to ensure a consistent comparison baseline. This was shown \textit{after} the initial questions to avoid bias. The final four questions focused on perceptions of the filler types and depended on participants’ awareness of them. While PSQ10 resembled PSQ4 and PSQ5 (future use intent), it was placed after the video to reflect informed responses.

\begin{table*}[ht]
\caption{Questions ordered as they appeared in the post-VR survey. Rank-order question order was randomly initialized. $\star$: text entry answer justification required; $\star\star$: optional.}
\label{tab:post_study}
\Description{Table lists the questions asked after participant finishes the in-VR portion of the study. Questions one through 6 (PSQ1 through PSQ6) are shown first. PSQ1 and PSQ2 ask which agent participant like the most and the least; PSQ3 allows the participant to agree or disagree with the statement "I feel like the agents understood me" on a scale from Strongly Agree (5) to Strongly Disagree (1); PSQ4 and PSQ5 use the same response anchors, asking users about future use intent, considering that virtual agents respond as fast as fastest agents responded in the study, and as slow as the slowest agents responded in the study; PSQ6 asks participants whether they have noticed any conversational fillers that different agents had, with options Yes, Maybe, No. Then, participants watched a video that explained the three filler types. Then, four more questions followed: PSQ7 asked participants to rank the conversational fillers in terms of their preference of seeing them in future applications; PSQ8 asked participants whether gestures or voice were more helpful in natural fillers; PSQ9 asked whether the visual wait indicator, or the sound effect was more helpful when artificial fillers were used; PSQ10 asked participants whether they would use a system where agents replied as slow as the slowest agents in the study, but considering that conversational fillers would be present, and this question used the same response anchors as PSQ4 and PSQ5.}
\begin{tabular}{p{.39in}p{4.8in}p{1.35in}}
\toprule
\textbf{Label} & \textbf{Question} & \textbf{Response anchors} \\
\midrule
\textbf{PSQ1} & Overall, which agent did you like the most? $\star$                                                                                    & Each of 9 agents                             \\ 
\textbf{PSQ2} & Overall, which agent did you like the least? $\star$                                                                                   & Each of 9 agents                             \\ 
\textbf{PSQ3} & I felt like the agents understood me. $\star$                                                                                         & Agree $\rightarrow$ Disagree           \\ 
\textbf{PSQ4} & I would use a system where agents replied fastest. $\star$                                                                           & Agree $\rightarrow$ Disagree           \\ 
\textbf{PSQ5} & I would use a system where agents replied slowest. $\star$                                                                            & Agree $\rightarrow$ Disagree           \\ 
\textbf{PSQ6} & Did you notice any conversational fillers that different agents had? $\star\star$                                                          & Yes, Maybe, No                               \\ 
\hline
\cellcolor{gray!20} & \cellcolor{gray!20} Video explanation of the three filler types. & \cellcolor{gray!20} \\
\hline
\textbf{PSQ7} & Rank the conversational fillers that you saw in terms of your preference for future applications.                              & Natural, Artificial, None \\ 
\textbf{PSQ8} & For \textbf{Natural} fillers, were \textit{gestures} or \textit{voice lines} more helpful to fill conversations? $\star\star$          & Gestures, Voice lines, Same        \\ 
\textbf{PSQ9} & For \textbf{Artificial} fillers, were \textit{wait indicators} or \textit{sound effects} more helpful to fill conversations? $\star\star$ & Indicators, Sounds, Same \\ 
\textbf{PSQ10} & I would use a system where agents replied slowest with conversational fillers. $\star$                                               & Agree $\rightarrow$ Disagree           \\
\bottomrule
\end{tabular}
\end{table*}

\subsection{Interaction Scenarios}

\begin{figure*}[!ht]
    \centering
    \includegraphics[width=.99\linewidth]{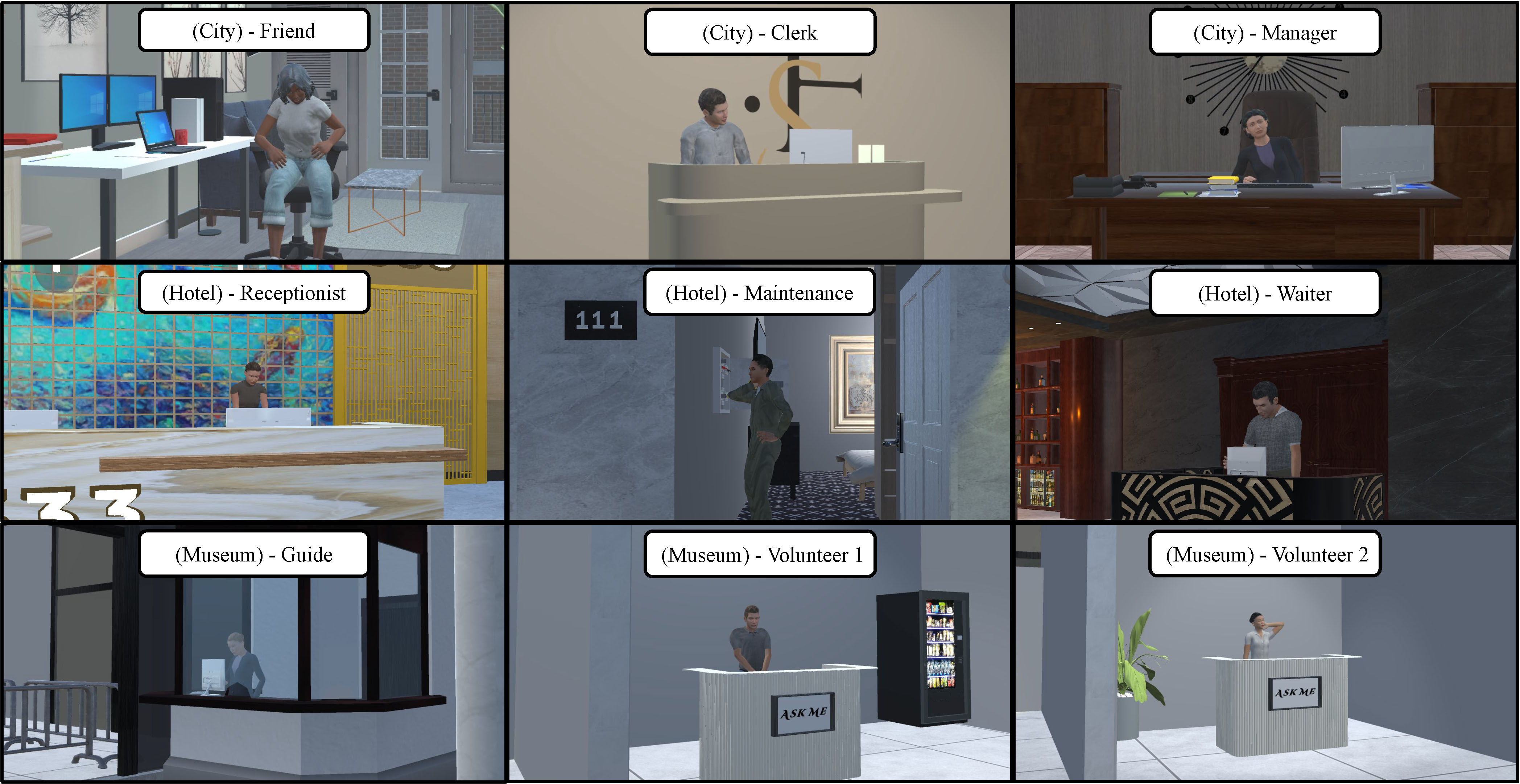}
    \caption{Intelligent Embodied Virtual Agents in their corresponding virtual environments.}
    \label{fig:agents_environments}
    \Description[Environments and Agents]{Figure shows the Intelligent Embodied Virtual Agents in their surrounding environments. The top left figure shows the entrance of the museum with the guide agent. The top middle figure shows the volunteer 1 within the museum standing behind a desk. The top right figure shows the volunteer 2 within the museum standing behind a desk. The middle left figure shows the hotel scene receptionist. The middle figure shows the maintenance worker fixing an electric board within a room. The middle right figure shows the hotel restaurant waiter. The bottom left figure shows the friend agent sitting on her desk chair. The bottom middle figure shows the clerk standing behind a desk. The bottom right figure shows the manager sitting behind her computer desk.}
\end{figure*}

Participants interacted with nine virtual agents across three environments (\autoref{fig:agents_environments}) corresponding to distinct scenarios: \textbf{Store}, \textbf{Hotel}, \textbf{Museum}. Each condition targeted at least five conversational turns to ensure the study conditions were noticeable~\cite{boukaram2021mitigating}, influencing participants' survey responses. To maintain engagement, the scenarios were designed with a gamified structure (similar to video game quests) and clear objectives, encouraging free-form conversations. Agents' responses were generated at runtime using LLM prompts (see project repository) and user queries, and supported small-talk (e.g., greetings, environment awareness). When queried with off-topic or role-breaking input, agents naturally redirected the conversation back to the scenario context. Participants were free to explore each environment and speak with agents, guided by an on-hand UI that updated after the transition-check system (\autoref{fig:typical_arch}) detected relevant dialogue context in the message history.

In the \textbf{Store} scenario, participants retrieved a shirt from the \textit{Friend} agent and returned it to a store across the street. There, the \textit{Clerk} directed them to the \textit{Manager} for approval, as he was still in training. 
In the \textbf{Hotel} scenario, participants checked in with the \textit{Receptionist}, then visited their room where \textit{Maintenance} worker informed them it wasn't ready. The worker apologized and redirected them back to the receptionist to receive a complimentary dinner voucher. At the restaurant, the \textit{Waiter} asked about food preferences, dietary restrictions, and the voucher. 
In the \textbf{Museum} scenario, participants played a student working on a school assignment about human rights. They obtained a ticket from the \textit{Host} agent, then spoke with \textit{Volunteer 1} about the Cyrus Cylinder\footnote{\href{https://en.wikipedia.org/wiki/Cyrus\_Cylinder}{en.wikipedia.org/wiki/Cyrus\_Cylinder}\accessdate} and \textit{Volunteer 2} about the U.S. civil rights movement.

\subsubsection{Intelligent Agents' Avatars}
\label{sec:agents}

Agents were represented using avatars that matched the visual fidelity of our virtual environments (VEs). \autoref{fig:agents_environments} shows the agents in context, with race and gender distributions reflecting our university's student demographics. While we initially considered Rocketbox avatars~\cite{rocketbox}, we selected avatars from the VALID library~\cite{doValid2023} due to their validation through user perception studies, ensuring accurate and representative character design. Agent outfits aligned with scenario roles and the visual style of each VE. Avatar folder names and corresponding Edge-TTS voices are listed in~\autoref{tab:agents}.





\begin{table}[ht]
\caption{Scenarios, roles, avatars, and abbreviated TTS voices used for avatars in our experiment. Full Edge-TTS voice identifiers are available in the source code repository.}
\label{tab:agents}
\Description{This table lists the virtual agents used in different interaction scenarios, including their roles, avatar names, and the corresponding text-to-speech voices from the Edge-TTS API. Full voice names used for speech synthesis are available in the source code repository.}
\begin{tabular}{llll}
\toprule
\textbf{Scenario} & \textbf{Role} & \textbf{VALID Avatar} & \textbf{Voice} \\
\midrule
Store & Friend & \texttt{Black\_F\_2\_Casual} & Ava \\
Store & Clerk & \texttt{White\_M\_3\_Casual} & Andrew \\
Store & Manager & \texttt{Hispanic\_F\_1\_Busi} & Aria \\
Hotel & Receptionist & \texttt{Hispanic\_F\_2\_Casual} & Michelle \\
Hotel & Maintenance & \texttt{Hispanic\_M\_2\_Util} & Guy \\
Hotel & Waiter & \texttt{MENA\_M\_2\_Casual} & Brian \\
Museum & Host & \texttt{White\_F\_2\_Busi} & Emma \\
Museum & Volunteer 1 & \texttt{White\_M\_1\_Casual} & Florian \\
Museum & Volunteer 2 & \texttt{Asian\_F\_1\_Casual} & Yan \\
\bottomrule
\end{tabular}
\end{table}

\subsubsection{Agent Animations}

Each agent performed a scenario-specific `busy' animation until first addressed by the user. For example, \textit{Waiter} and \textit{Host} interacted with small displays, \textit{Maintenance} worker manipulated wires on an electrical panel, and the \textit{Receptionist} and \textit{Manager} alternated between typing and looking at a monitor. Upon user interaction, agents transitioned to an `attentive idle' state (passive listening pose with subtle breathing). While users spoke, agents turned their heads toward them. Interactions were only possible within a defined proximity; agents would look away and return to their `busy' animation once the user exited this area.

\subsection{Conversational System Implementation}

\begin{figure*}[ht]
    \centering
    \includegraphics[width=.99\textwidth]{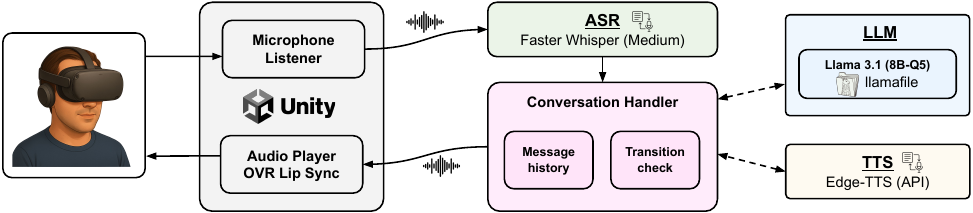}
    \caption{System architecture: speech is recorded in Unity, passed to the ASR model, combined with the message history, passed to an LLM, the generated response from which updates message history and checks for transitions. Generated text is then passed to TTS, and the generated voice is played from an audio source that controls OVR lip-sync in Unity.}
    \label{fig:typical_arch}
    \Description{System diagram shows a flowchart of how data was passed between multiple systems to achieve the real-time conversation functionality. User's voice is recorded with a Unity application in VR and passed to an automatic speech recognition system. The speech recognition system passes the transcribed text to a "Conversation Handler", which does two things: updates the message history for the agent that user spoke to, and also checks whether a transition to the next phase of the study should occur. The text response from the perspective of the agent is then generated using a large language model, specifically Llama version 3.1. This text response is then passed to a speech synthesis API named Edge text-to-speech, and the audio is sent back to the Unity application for to be played for the use.} 
\end{figure*}

\begin{figure}
    \centering
    \includegraphics[width=0.99\columnwidth]{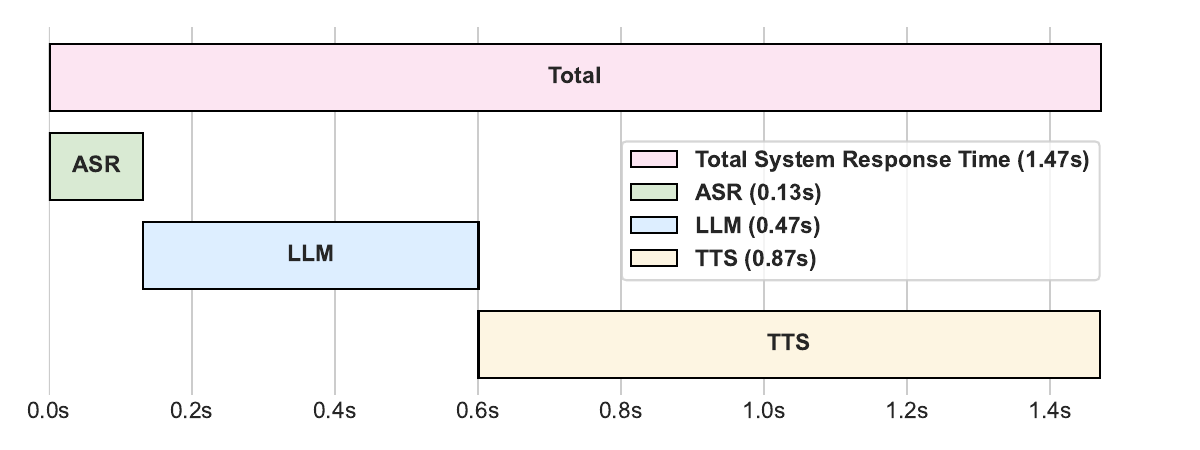}
    \caption{Breakdown of latencies achieved in our system. Latency is the time between the user finishing their microphone input (by pressing the `A' button) and when the agent starts responding with the LLM-generated text in voice.}
    \label{fig:local_latency}
    \Description[Latency conditions]{Breakdown of latencies achieved in our system. Latency is the time between the user finishing their mic input (by pressing the `A' button) and when the agent starts responding with the LLM-generated text in voice. Total system response time is 1.47 seconds, out of which ASR takes 0.11 seconds, LLM takes 0.49 seconds, and TTS takes 0.88 seconds.} 
\end{figure}

Our system (see~\autoref{fig:typical_arch}) was implemented as a Unity application connected via HTTP requests to a local server that served the Automatic Speech Recognition (ASR) and LLM engines. When a user spoke within the agent's defined area, the user's voice was transcribed with the FasterWhisper\footnote{\href{https://github.com/SYSTRAN/faster-whisper}{github.com/SYSTRAN/faster-whisper}\accessdate} medium model, then added to the message history of that agent, and then sent to the locally-hosted Llama3.1-8b-Q5 LLM\footnote{\href{https://huggingface.co/bullerwins/Meta-Llama-3.1-8B-Instruct-GGUF/tree/828492ca0d7e7efd4b316e75af8d9cd582fdec34}{huggingface.co/bullerwins/Meta-Llama-3.1-8B-Instruct-GGUF} Llama 3.1 achieves state of the art performance on a number of benchmarks at the time of our research\accessdate} to generate a response from the agent's perspective (ran on an RTX4090 GPU on Pop!\_OS22\footnote{\href{https://pop.system76.com/}{pop.system76.com}\accessdate}). This text was then passed to the Edge-TTS API\footnote{\href{https://github.com/rany2/edge-tts}{github.com/rany2/edge-tts}\accessdate}, which generated an audio file, storing it on our local server, and returning a static download link to Unity. The audio was then downloaded by the application and played from an audio source on the agent, with OVR Lipsync\footnote{\href{https://developers.meta.com/horizon/documentation/unity/audio-ovrlipsync-unity/}{developers.meta.com/horizon/documentation/unity/audio-ovrlipsync-unity/}\accessdate} used to animate the agent's mouth movements. The average system response time (SRT) was approximately 1.5 seconds ($\mu = 1.47$, $\sigma = .23$), constrained by the processing time of the locally hosted ASR, LLM, and TTS components (see \autoref{fig:local_latency} for a breakdown). This value defined the \LowLatency{} condition in our experiment.


\subsection{Apparatus}
We used a Meta Quest Pro HMD, with a resolution of $1800 \times 1920$ pixels per eye and a FOV of $106^{\circ} \times 96^{\circ}$  connected to a PC running the Unity 2022 LTS application. We developed the VR interactions using the XR Interaction Toolkit package\footnote{\href{https://docs.unity3d.com/Packages/com.unity.xr.interaction.toolkit@2.5/manual/index.html}{Unity Docs | XR Interaction Toolkit}\accessdate}. Our application ran at constant 70 frames per second (FPS). Participants wore sanitized earphones, and navigated the VE using the Quest Touch Pro controllers. The thumbsticks controlled movement and turning, the back trigger selected in-VR survey responses, the grip button selected objects within the VE, and the `A' button on the right controller activated the microphone.

\subsection{Participants}


We used G*Power~\cite{faul2009statistical} to estimate a minimum required sample size of 33 participants, assuming a medium-to-large effect size ($0.3$), a within-subjects design, and repeated measures ANOVA with 9 measurements. To increase statistical power and accommodate counterbalancing, our final participant pool included 54 university participants (three full rotations of the 18-order Latin square), comprising 23 females ($43\%$) and 31 males ($57\%$), aged 18-56 ($\bar{x} = 22.07$,~$s = 6.55$). Participants self-reported varying experiences:
\begin{itemize}[left=0pt, labelsep=0.5em, itemsep=0pt, parsep=0pt]
    \item \textbf{VR use:} Daily:~2, Weekly:~6, Monthly:~11, Yearly:~18, Never:~17
    \item \textbf{Gaming:} Daily:~16, Weekly:~16, Monthly:~12, Yearly:~7, Never:~3
    \item \textbf{Social VR\footnote{VRChat (\href{https://vrchat.com/}{vrchat.com}), Meta Horizon (\href{https://horizon.meta.com/}{horizon.meta.com})\accessdate}:} Daily:~0, Weekly:~2, Monthly:~13, Yearly:~8, Never:~31
\end{itemize}
All participants could read and speak English, wore an HMD for 35 minutes while seated, used both hands for controllers, and had normal or corrected vision. Those with glasses or contact lenses kept them on during the study. No participants reported color blindness, neurological conditions, or physical disabilities.

\subsection{Procedure}

Participants arrived at the study location and were screened for eligibility. After confirming eligibility, the consent process was administered, including answering any participant questions. Participants were then asked to complete a demographics survey electronically. Following this, participants were assisted in wearing the VR headset while seated. We adjusted the participants' seated in-VR height to their real-world standing height to make their conversations with the agents feel more natural. Before starting each scenario, we presented participants with a brief introduction, outlining the scenario's theme without revealing the outcome. In each scenario, participants had to navigate the environment and engage in conversations with three agents, totaling nine virtual agents across all scenarios (see~\autoref{sec:experiment_design}). To facilitate the exploring of the environment, participants had access to an on-hand task list attached to their left hand. This list only gave participants an overview of their active task without providing hints about the conversation flow. Once completed, a new task appeared, while the previous task remained as strike-through. Upon completing the 3 in-VR scenarios, participants removed the HMD and filled out a web-based post-study survey. We paid participants \$10 and thanked them for their time.

\section{Results}

\subsection{Post-condition Responses}

\begin{figure*}[bp]
    \centering
    \includegraphics[width=.99\linewidth]{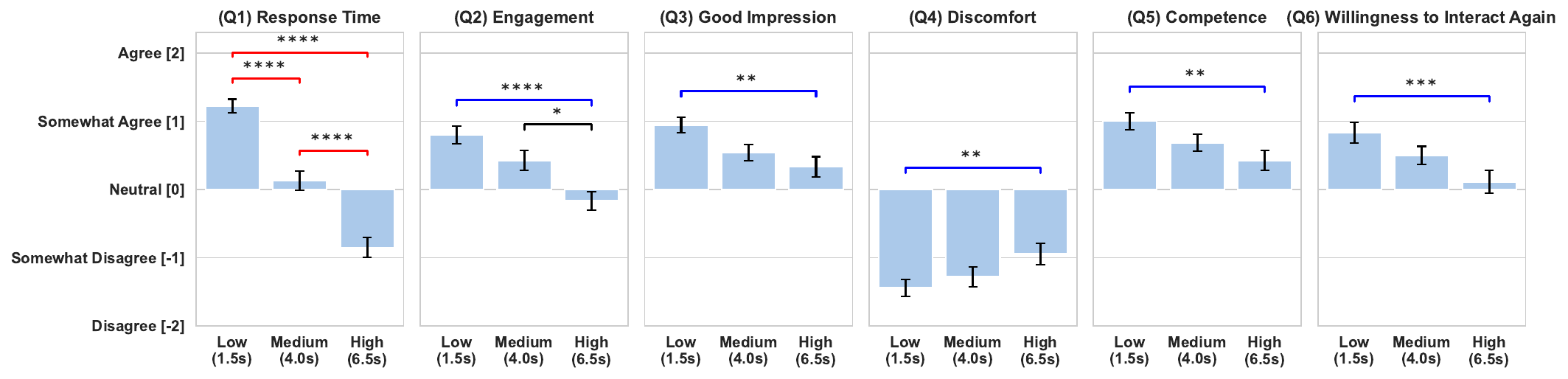}
    \caption{Effect of response delay on user perception of intelligent embodied virtual agents in the \textit{None} filler condition, collected through post-condition survey responses. {\color{red}\textbf{H1a}} -- Significant effect of delay on (Q1) \PerceivedLatency; {\color{blue}\textbf{H1b}} -- Significant effect of delay on other perception dimensions. Error bars show 1 SEM (standard error of the mean). Significance annotations~\cite{florian_charlier_2022_7213391} from ART-C pairwise comparisons adjusted using Holm-Bonferroni (* = $p<0.05$, ** = $p<0.01$, *** = $p<0.001$, **** = $p<0.0001$).}
    \label{fig:H1}
    \Description{The figure shows that all dimensions of participants' perception of the agents were significantly affected by latency, when no fillers were present. Perceived response time, engagement, good impression, competence, and willingness to interact with agents again decreased with the increase of delay. Discomfort also increased with the increase in response delay. The figure shows a significant difference between Low (1.5 seconds) and High (6.5 seconds) latencies across all questions collected in the post-condition surveys. More details of the Figure are described in the results section.}
\end{figure*}

\begin{figure*}
    \centering
    \includegraphics[width=0.99\linewidth]{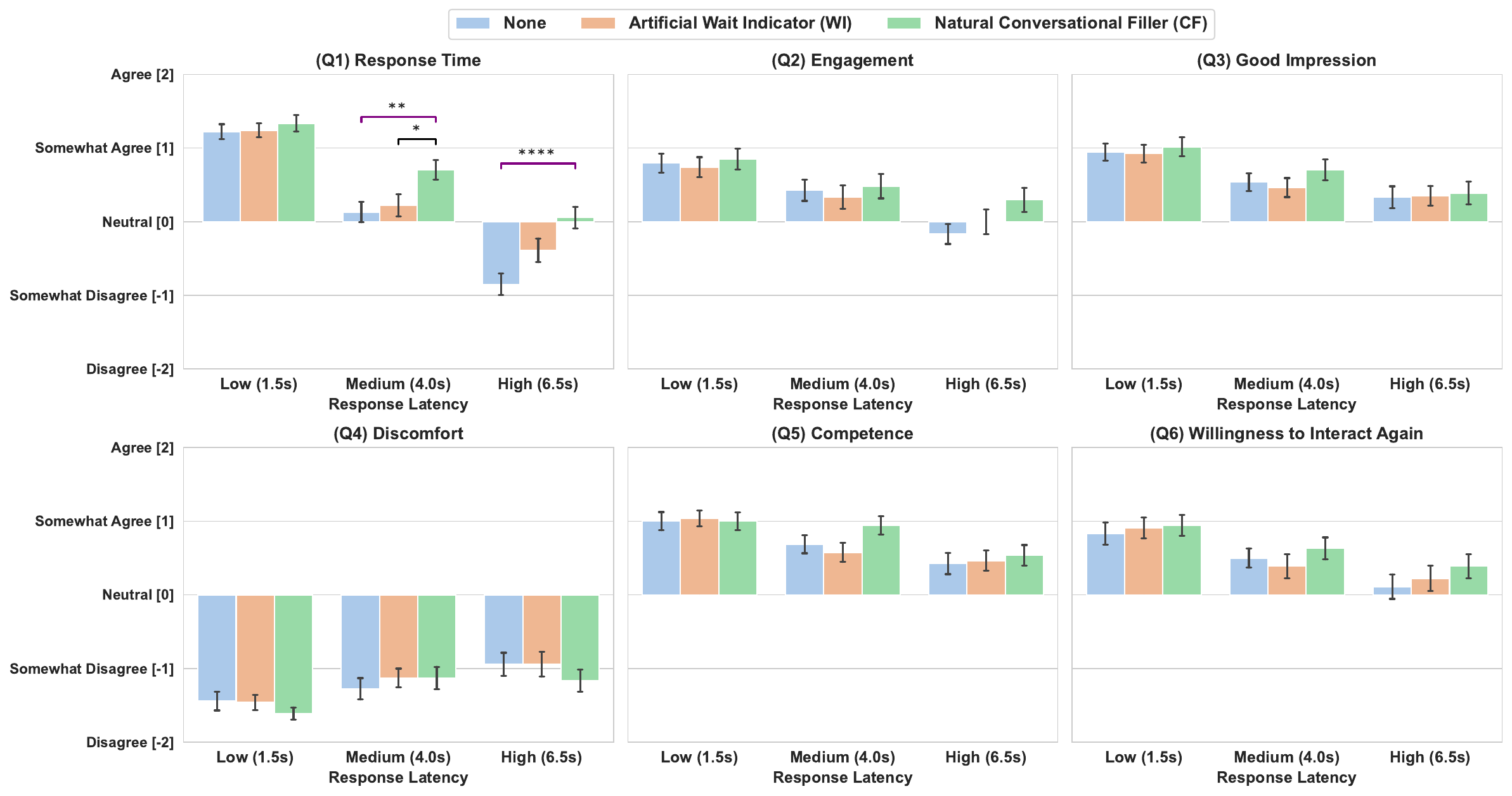}
    \caption{Effect of delay with all filler types on user perception of intelligent embodied virtual agents. {\color{violet}\textbf{H2a}} -- Significant effect of \textit{Natural Conversational Fillers} on (Q1) \PerceivedLatency; Error bars show 1 SEM (standard error of the mean). Significance annotations~\cite{florian_charlier_2022_7213391} from ART-C pairwise comparisons adjusted using Holm-Bonferroni (* = $p<0.05$, ** = $p<0.01$, *** = $p<0.001$).}
    \label{fig:H2}
    \Description{Figure shows the effect of delay under the presence of different filler types, and how it affected participant's responses to the post-condition questionnaires. Participants responses trended with the data presented in the no filler type, however, Natural fillers significantly improved participants' for the (Q1) Response Time responses. Artificial fillers were not effective at improving the perception dimensions. Details of the Figure are also described in the results section.}
\end{figure*}

Since our data consists of Likert-scale responses, the Aligned Rank Transform (ART)~\cite{wobbrock2011aligned} was used for a 3×3 full-factorial repeated-measures ANOVA. This analysis examined main and interaction effects across three delay levels (\LowLatency, \MediumLatency, and \HighLatency) and three filler types (\textit{None}, \textit{Artificial Wait Indicator}, and \textit{Natural Conversational Filler}). Post-hoc ART-C tests~\cite{artc} were conducted to test the hypotheses, with p-values adjusted for 36 comparisons using the Holm-Bonferroni correction. Responses were coded from $-2$ (Disagree) to $2$ (Agree), and error bars in \autoref{fig:H1} and \autoref{fig:H2} show $\pm$1 SEM based on this scale.

\begin{table}[ht]
\caption{Repeated-Measures ANOVA on ART-Transformed Data. $\eta_p^2$ is the effect size of a specific independent variable on the dependent variable (metric measured by the question), while controlling for effects of other independent variables.}
\label{tab:anova}
\Description{The table shows the results of the repeated measures ANOVA test, applied to six post-condition questions, to test for significant main and interaction effects of the Delay (latency level) and the Filler type. For Delay, large effect size was detected for (Q1) Response Time; medium effect size was detected for (Q2) Engagement, (Q3) Good Impression, (Q5) Competence, and (Q6) Willingness to interact again; small effect size was detected for (Q4) Discomfort. Filler type had a medium effect size on (Q1) Response Time; small effect size of Filler type was also detected for (Q2) Engagement, (Q3) Good Impression, and (Q5) Competence. Small interaction effect Delay and Filler was only detected for (Q1) Response time.}
\scriptsize
\begin{tabular}{|l|l|c|c|c|c|c|}
\hline
\multicolumn{1}{|c|}{\textbf{Question}} & \multicolumn{1}{c|}{\textbf{Factor}} & \textit{\textbf{Df}} & \textit{\textbf{F}} & \textit{\textbf{p}} & \textit{\textbf{Sig.}} & \textit{\textbf{$\eta_p^2$}} \\ \hline
\multirow{3}{*}{\textbf{(Q1)} Response Time} & Delay & $(2, 424)$ & $154.55$ & \textit{p} < 0.0001 & **** & $0.422$ \\
 & Filler & $(2, 424)$ & $18.61$ & \textit{p} < 0.0001 & **** & $0.081$ \\
 & Delay x Filler & $(4, 424)$ & $4.66$ & \textit{p} < 0.01 & ** & $0.042$ \\
\hline
\multirow{3}{*}{\textbf{(Q2)} Engagement} & Delay & $(2, 424)$ & $25.33$ & \textit{p} < 0.0001 & **** & $0.107$ \\
 & Filler & $(2, 424)$ & $4.68$ & \textit{p} < 0.01 & ** & $0.022$ \\
 & Delay x Filler & $(4, 424)$ & $1.08$ & \textit{p} = 0.365 &  & $0.010$ \\
\hline
\multirow{3}{*}{\textbf{(Q3)} \makecell[l]{Good\\Impression}} & Delay & $(2, 424)$ & $30.92$ & \textit{p} < 0.0001 & **** & $0.127$ \\
 & Filler & $(2, 424)$ & $5.03$ & \textit{p} < 0.01 & ** & $0.023$ \\
 & Delay x Filler & $(4, 424)$ & $0.36$ & \textit{p} = 0.834 &  & $0.003$ \\
\hline
\multirow{3}{*}{\textbf{(Q4)} Discomfort} & Delay & $(2, 424)$ & $4.65$ & \textit{p} < 0.05 & * & $0.021$ \\
 & Filler & $(2, 424)$ & $1.28$ & \textit{p} = 0.280 &  & $0.006$ \\
 & Delay x Filler & $(4, 424)$ & $1.55$ & \textit{p} = 0.187 &  & $0.014$ \\
\hline
\multirow{3}{*}{\textbf{(Q5)} Competence} & Delay & $(2, 424)$ & $14.74$ & \textit{p} < 0.0001 & **** & $0.065$ \\
 & Filler & $(2, 424)$ & $3.74$ & \textit{p} < 0.05 & * & $0.017$ \\
 & Delay x Filler & $(4, 424)$ & $1.18$ & \textit{p} = 0.320 &  & $0.011$ \\
\hline
\multirow{3}{*}{\textbf{(Q6)} \makecell[l]{Willingness to\\Interact Again}} & Delay & $(2, 424)$ & $16.93$ & \textit{p} < 0.0001 & **** & $0.074$ \\
 & Filler & $(2, 424)$ & $2.26$ & \textit{p} = 0.106 &  & $0.011$ \\
 & Delay x Filler & $(4, 424)$ & $0.42$ & \textit{p} = 0.791 &  & $0.004$ \\
\hline

\end{tabular}

\end{table}

Delay had a significant main effect on (Q1) \PerceivedLatency, (Q2) Engagement, (Q3) Good Impression, (Q5) Competence, and (Q6) Willingness to Interact Again (all $p<0.0001$). It also significantly affected (Q4) Discomfort ($p<0.05$). The main effect of fillers was significant for (Q1) \PerceivedLatency{} ($p<0.0001$), (Q2) Engagement ($p<0.01$), (Q3) Good Impression ($p<0.01$), and (Q5) Competence ($p<0.05$). Interactions between latency levels and Fillers were significant only for (Q1) \PerceivedLatency{} ($p<0.01$).

Post-hoc pairwise comparisons between delay levels in the absence of fillers (\autoref{fig:H1}) revealed significant differences in (Q1) \PerceivedLatency{} across all conditions ($p<0.0001$), supporting \textbf{H1a}. For other metrics, participants’ responses significantly differed between \LowLatency{} and \HighLatency{} delay levels on (Q2) Engagement ($p<0.0001$), (Q3) Good Impression ($p<0.01$), (Q4) Discomfort ($p<0.01$), (Q5) Competence ($p<0.01$), and (Q6) Willingness to Interact Again ($p<0.001$), supporting \textbf{H1b}. Additionally, (Q2) Engagement differed significantly between \MediumLatency{} and \HighLatency{} delays ($p<0.05$). 

Pairwise comparisons between filler types at fixed delay levels (\autoref{fig:H2}) showed significant differences between \textit{Natural} and \textit{None} fillers on (Q1) \PerceivedLatency{} at \MediumLatency{} ($p<0.01$) and \HighLatency{} ($p<0.0001$) delays, supporting \textbf{H2a}. \textit{Natural} and \textit{Artificial} fillers also significantly differed at \HighLatency{} delay on (Q1) \PerceivedLatency{} ($p<0.05$). However, no support was found for \textbf{H2b}, \textbf{H3a}, or \textbf{H3b}. We discuss implications of these results in~\autoref{sec:discussion_hypotheses1}, and a full report of all main effects and pairwise comparisons is available in the supplementary material.

\subsection{Post-study Responses}

\subsubsection{Preference for agents depending on response speed}
The first two post-study questions assessed whether participants preferred agents that responded quickly. The frequency of each speed condition was recorded for agent selections. Among the agents participants liked the most (PSQ1), \LowLatency{} responses occurred 31 times (57.41\%), \MediumLatency{} 12 times (22.22\%), and \HighLatency{} 11 times (20.37\%), ($\chi_{2}^2 (N=54) = 14.12, p < 0.0001$). Among the agents participants liked the least (PSQ2), \LowLatency{} responses occurred 9 times (16.67\%), \MediumLatency{} 12 times , and \HighLatency{} 33 times (61.11\%), ($\chi_{2}^2 (N=54) = 19.0, p < 0.0001$).

\subsubsection{Agents' understanding and noticeability of conversational fillers}
Overall, 49 participants (90.74\%) indicated that the agents understood them through answers to PSQ3 (\textit{Strongly Agree}: 22, \textit{Agree}: 27, \textit{Neither agree nor disagree}: 3, \textit{Disagree}: 2, \textit{Strongly disagree}: 0). A chi-square test revealed a significant difference from a uniform choice distribution ($\chi_{4}^2 (N=54) = 36.82, p <  0.0001$). Before being introduced to the filler types through a video, most participants reported noticing them on PSQ6 (\textit{Yes}: 33, \textit{Maybe}: 11, \textit{No}: 10), with non-uniform responses distribution ($\chi_{2}^2 (N=54) = 18.78, p < 0.0001$).

\subsubsection{Preference for filler types in future applications}
\begin{figure}
    \centering
    \includegraphics[width=0.99\columnwidth]{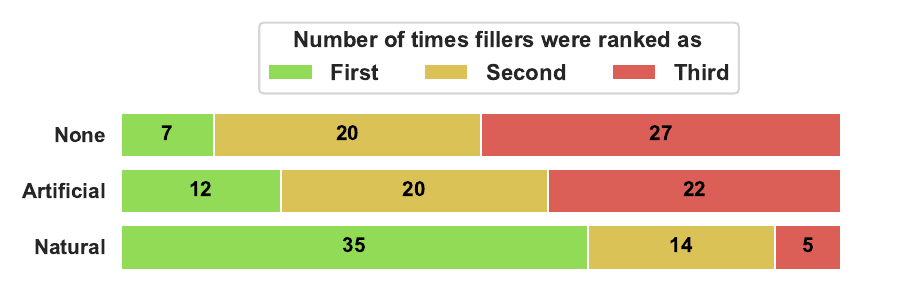}
    \caption{Number of times each filler was ranked in terms of preference of seeing it in future applications.}
    \Description{Number of times each filler was ranked in terms of preference of seeing it in future applications, showing that Natural fillers were ranked as a top choice by 33 participants out of 54, and second by 14 participants out of 54. It also shows that "no fillers" were ranked lowest -- only selected by 7 participants as their top choice, and 27 selected it as their least preferred delay mitigation strategy. Artificial fillers were selected by 12 as their to pick, by 20 as their second pick, and by 22 as their third pick. The data from this figure is also described in text.}
    \label{fig:filler_rankings}
\end{figure}

In PSQ7, Participants ranked the three types of fillers based on their preference for their use in future applications. \autoref{fig:filler_rankings} shows that the majority of participants (35/54, 64.81\%) selected \textit{Natural} filler as the most preferred, while 12 participants (22.22\%) favored the \textit{Artificial} fillers instead. Seven (12.96\%) participants chose \textit{None} (no filler) as their top preference.

\subsubsection{Helpfulness of visual and auditory modalities of fillers}
\label{sec:modality_importance}
For \textit{Natural} fillers (PSQ8), 16 participants (29.70\%) found gestures more helpful, 14 (25.93\%) chose voice lines, and 24 (44.44\%) rated both equally ($\chi_{2}^2 (N=54) = 3.12, p = 0.21$). For \textit{Artificial} fillers (PSQ9), 19 participants (35.18\%) preferred the visual wait indicator, 10 (18.52\%) preferred sound effects, and 25 (46.30\%) reported that both contributed equally ($\chi_{2}^2 (N=54) = 6.34, p < 0.05$). 

\subsubsection{Statements about using the system depending on the speed of agent's responses}
\begin{figure*}
    \centering
    \includegraphics[width=0.99\textwidth]{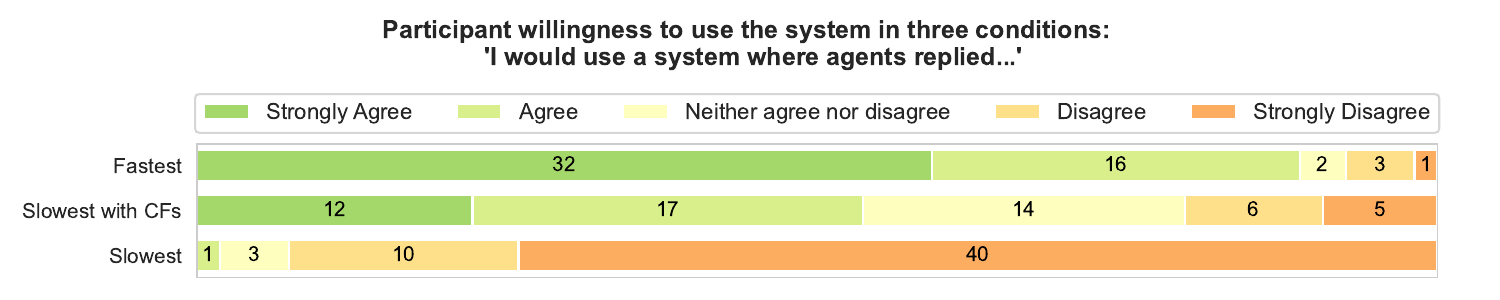}
    \caption{Participants' responses to question on whether they would use the system depending on the speed of agents' responses (PSQ4, PSQ5) and presence of conversational fillers (PSQ10).}
    \label{fig:use_speeds}
    \Description{Figure shows the distribution of participants' responses on question whether they would use the system depending on the speed of agents' responses and presence of conversational fillers (CFs). It highlights how most participants strongly agreed that they would use the system if the agents responded fastest, and strongly disagreed (40 out of 54) that they would use a system where agents responded slowest. It also shows that participant's responses significantly shifted from strongly disagreeing to agreeing that they would use the slowest system, if the system contained conversational fillers. Data from this figure is also described in Results and analyzed in Discussion.}
\end{figure*}

In PSQ4, PSQ5 and PSQ10, participants responded to whether they would use the system with agents at different response latencies and with or without conversational fillers (see~\autoref{fig:use_speeds}). A chi-square test revealed that the distribution of choices for the use of the fastest agents ($\chi_{4}^2 (N=54) = 65.81, p < 0.001$), slowest agents ($\chi_{4}^2 (N=54) = 104.33, p < 0.001$), and slowest agents with conversational fillers ($\chi_{4}^2 (N=54) = 9.89, p < 0.05$) were not uniform. A comparison of participants' willingness to use the slowest system with and without conversational fillers revealed greater acceptance when fillers were present ($\chi_{4}^2 (N=54) = 61.56, p < 0.001$).

\subsubsection{Analysis of text justifications}

To analyze participants' text justifications of their post-study questionnaire answers,
we conducted a preliminary inductive analysis using in vivo excerpts and descriptive coding to generate general thematic insights~\cite{bingham_data_2023}, and counted the frequencies of similar statements across participants.

\section{Discussion}

To our knowledge, our study is the first to vary both response latency and conversational fillers in an experiment where conversational virtual agents were \textit{truly interactive}, leveraging ASR, LLM and TTS to support free-form conversations. Previous research on latency mitigation has largely studied it in isolation, rather than including it as an explicit factor in studies with fully interactive embodied conversational agents. We found that response latency significantly degrades user perceptions of the agents, and that conversational fillers improve perceived response latency in \MediumLatency{} and \HighLatency{} delay levels. These findings align with prior work that investigated conversational fillers in WoZ studies~\cite{wigdor2016improve, funakoshi_smoothing_2008, kum_can_2022, boukaram2021mitigating, jeong2019exploring, shiwa_how_2009} and those that used pre-recorded human-agent conversations~\cite{ohshima_conversational_2015, lopez2019testing}. This section addresses our research questions by interpreting participants' post-condition responses to evaluate hypotheses (\autoref{sec:discussion_hypotheses1}, \autoref{sec:discussion_hypotheses2}). We then interpret the remaining data to derive insights about acceptable response latency in~\autoref{sec:acceptable_latency}, effect of human-likeness of agents in~\autoref{sec:human_likeness}, visual and auditory modalities in~\autoref{sec:modalities}, and implications on future work in~\autoref{sec:implications}, informing future studies and the design of human-agent conversational user interfaces.

\subsection{Effects of Response Latency}
\label{sec:discussion_hypotheses1}

Response latency had a significant negative impact on perceived (Q1) \PerceivedLatency{} (\autoref{fig:H1}), with all pairwise comparisons between delay conditions showing significant differences ($p<0.0001$), confirming \textbf{H1a}. This aligns with prior research on response delays in human-computer interaction~\cite{wigdor2016improve, funakoshi_smoothing_2008, kum_can_2022, boukaram2021mitigating, jeong2019exploring, ohshima_conversational_2015, lopez2019testing, shiwa_how_2009}, reinforcing that conversational latency directly impacts perceived system efficiency. Additionally, our results support \textbf{H1b} --- higher response latency was associated with lower (Q2) Engagement, (Q3) Good Impression, (Q5) Competence, (Q6) Willingness to Interact Again, as well as increased (Q4) Discomfort. These findings suggest that beyond perceived response time, delays in responses affect user trust and willingness to continue interacting with virtual agents, which is consistent with earlier work on conversational interruptions and delay tolerance~\cite{kum_can_2022}.

Post-study responses to PSQ1 and PSQ2 revealed that most participants favored agents that corresponded to the \LowLatency{} latency (31/54, 57.41\%), and disliked agents that corresponded to \HighLatency{} latency (33/54, 61.11\%). Given that Delay-Filler conditions applied to agents were counterbalanced between participants (implying an equal distribution (18/54, 33.33\%) per delay level if chosen randomly), these preferences suggest that latency shaped participants' impressions of agents, even though they were not explicitly informed about response speeds. In text justifications for these two questions, 25/54 (46.29\%) participants directly mentioned fast response times as the main reason behind choosing their favorite agent, and 22/54 (40.74\%) participants mentioned slow response times as the reason for choosing an agent as their least favorite. Results from PSQ4 and PSQ5 (\autoref{fig:use_speeds}) suggest that response speed was a key determinant in participants' system preference. Most participants indicated that they would use a system with the fastest response times, but rejected systems that responded slowest, which aligns with literature indicating that increased delay reduces future use intent~\cite{hoxmeier2000system}. These findings should be further validated through future studies involving conversational systems, through questionnaires containing concrete latency reference points and use case scenarios.

\subsection{Effects of Natural and Artificial Fillers}
\label{sec:discussion_hypotheses2}

\textit{Natural} conversational fillers significantly improved participants' ratings on (Q1) \PerceivedLatency{} at \MediumLatency{} and \HighLatency{} latency levels ($p<0.01, p<0.0001$, respectively), supporting \textbf{H2a}. This finding is consistent with prior studies indicating that conversational fillers facilitate smoother interactions and reduce the cognitive burden of waiting~\cite{boukaram2021mitigating, kum_can_2022}. Although \textit{Natural} fillers improved average ratings on broader user experience dimensions over no fillers (\textit{None} filler condition), enhancements in Engagement, Good Impression, Discomfort, Competence, or Willingness to Interact Again were not significant. Therefore, \textbf{H2b} was not supported, suggesting that while \textit{Natural} conversational fillers improve perceived response latency, they do not mitigate its broader effects on user experience.

Although average \PerceivedLatency{} ratings were slightly higher with \textit{Artificial} fillers at \MediumLatency{} and \HighLatency{} latency levels, they were not significantly different from the \textit{None} condition (see \autoref{fig:H2}). This result does not support \textbf{H3a}. Similarly, participants' ratings on broader  perception dimensions were not significantly different from \textit{None} fillers, so \textbf{H3b} was also not supported. These findings suggest that passive visual and auditory wait indicators (e.g., visual loading icons and sounds) are not engaging enough to mitigate response latency or improve user experience on broader dimensions.

Responses to PSQ7 showed that \textit{Natural} fillers ranked highest among the participants (35/54, 64.81\%), as compared to \textit{Artificial} fillers and no fillers present (see~\autoref{fig:filler_rankings}). Most participants also rejected the idea of using a system where agents responded slowest (PSQ5). However, they were more open to such system when conversational fillers were present (PSQ10), with a statistically significant shift in responses ($\chi_{4}^2 (N=54) = 61.56, p < 0.001$). While this suggests that fillers can improve tolerance for delayed responses, participants still showed an overall preference for faster responses.

\subsection{Insights on Acceptable Response Latency}
\label{sec:acceptable_latency}

Through our study, we empirically demonstrated that \textbf{response latency above 4 seconds significantly degrades user experience} in conversations with intelligent embodied virtual agents: 
(1) perceived latency degraded at \MediumLatency{} and \HighLatency{} latencies (\autoref{fig:H1}); 
(2) future use intent was lowest under high-delay conditions (\autoref{fig:use_speeds});
(3) agents with slowest response times were disproportionally ranked lowest in the post-study survey (\autoref{sec:discussion_hypotheses1}). These findings are especially relevant amid the growing adoption of conversational user interfaces powered by ASR, LLM, and TTS pipelines. Recent studies using similar pipelines for free-form interaction frequently omit explicit reporting of system response speed, despite descriptions suggesting relatively high latencies~\cite{garcia_speaking_2024, zhu_free_form_2023, yang_effects_2024}. Based on our results, we recommend that future studies minimize and mitigate response delays, aiming for latencies under 4 seconds. Failing to do so risks negatively biasing participants’ perceptions of conversational agents and degrading overall system usability.

While incremental response generation techniques~\cite{tsai2019faster, skantze2013towards} can improve perceived response time, they are not always viable --- critical information at the end of a user’s turn may require a complete restructuring of the generated response. LLM-to-TTS transfer, where voice generation can begin as soon as the first sentence is produced, is the most parallelizable component in recent pipelines. Advancements, such as OpenAI’s GPT-4o\footnote{\href{https://openai.com/index/hello-gpt-4o/}{openai.com/index/hello-gpt-4o}\accessdate}, demonstrate the potential for responses with negligible latency by processing input and output directly in speech form, bypassing sequential steps that involve text. However, these advancements do not address latency introduced by network instability or hardware limitations, which persistently cause latency in real-world deployments.

Despite ongoing efforts to accelerate response generation through model optimization and hardware improvements, the demand for more advanced reasoning capabilities inherently increases computational overhead. Techniques such as retrieval-augmented generation (RAG) and web search integration improve factual accuracy by dynamically incorporating external knowledge, but at the cost of additional processing time. Similarly, chain-of-thought (CoT) reasoning enhances logical inference by explicitly generating multi-step responses, while test-time optimization (TTO) adapts model outputs based on recent interactions --- both substantially increasing SRT. These reasoning techniques are crucial in areas where accuracy outweighs speed, such as medical, legal, financial, and educational domains.
However, their computational costs underscore a repeating pattern: as systems become faster, expectations rise, and increasingly complex compute causes latency to reemerge. Thus, mitigating IVA response latency at the user interface level --- through adaptive UI design and conversational fillers --- is just as critical as optimizing SRT. Rather than treating response delay as a purely technical limitation, future work should treat latency as an inevitable factor in high-quality conversational AI and design interactions that minimize its negative perceptual impact.

\subsection{Participants' Preference for Human-likeness of Avatars}
\label{sec:human_likeness}

When designing our study, we focused on making the agents appear, behave, and sound human-like (natural). For this reason, we used VALID avatars (see~\autoref{sec:agents}), animated all agents with `busy', `attentive idle', and `thinking' states, as well as used a TTS engine that generated realistic voices for responses. In addition, conditions with \textit{Natural} fillers included a thinking gesture and a voice line while agents generated their responses. In justifications for choosing their favorite agent, a majority of participants ($38/54, 70.37\%$) included naturalness and relatability of conversations, in addition to quick responses. Some participant quotes provide more insight: 
(1) "\textit{She \textbf{responded surprisingly quickly} and her \textbf{answers felt lifelike} compared to most of the other AIs I interacted with}",
(2) "\textit{The agent was very \textbf{personable} and acted like an actual friend would. The most \textbf{natural} of all of the agents}". 
Conversely, participants who selected specific agents as their least favorite often cited issues related to the lack of human-likeness: unnaturalness ($20/54, 37.03\%$), awkwardness ($11/54, 20.37\%$), and lack of motion during wait time ($20/54, 37.03\%$). For example: 
(1) "\textit{Felt the most \textbf{robot like} of the agents, with \textbf{delayed responses} and the dialogue did not feel smooth}",
(2) "\textit{It was fast but \textbf{creepy}}". 

While participants favored human-like avatars, presence of \textit{Natural} fillers at \LowLatency{} latency did not significantly change perception on neither (Q1) Response Time, nor the broader dimensions of perception of the agents. This suggests that when agents are quick to respond, designers should prioritize delay being filled with animations consistent with agent's behavior, as execution of \textit{Natural} fillers in a short time frame could be perceived as rushed and exaggerated. Moreover, in medical and other high-stakes scenarios, human tendency to trust and be more forgiving of anthropomorphized AI~\cite{weiz1966eliza, huang_human_2025, hu_is_2024} may discourage users from scrutinizing accuracy. Future work should further investigate the relationship between filler anthropomorphism and risk profile: richer, natural cues for social or entertainment settings; more neutral or artificial signals where critical judgment matters.

\subsection{Relative Importance of Visual and Audio Modalities}
\label{sec:modalities}

Responses on the different modalities of the \textit{Natural} and \textit{Artificial} fillers revealed that many participants felt they contributed similarly (see~\autoref{sec:modality_importance}). However, quotes from participants with opposing opinions provide interesting insights. For \textit{Natural} fillers, quotes in favor of gestures include: (1) "\textit{Gestures are a normal motion people do when prompted with questions or problems. So seeing the agents do it made it more of a real world experience}.", (2) "\textit{It made me subconsciously realize they were thinking. But I preferred when they made a short hmm and a gesture}". On the other hand, some participants favored voice fillers over gestures, reasoning about higher naturalness, giving agents a buffer time to think, and confirming that the agents heard them; sample quotes include: (1) "\textit{It seemed unrealistic and almost cartoony that the AI would go into a thinking pose every time I said something to them. The filler voice line, however, was surprising at first, though it felt like a natural buffer to give them time to think that you would see in a regular person.}", and (2) "\textit{It let me know that the agent heard me. If they just used gestures, I would assume it's an idle animation.}".

For \textit{Artificial} filler conditions, participants favored visual cues because of familiarity, for example: "\textit{I'm used to the loading UI because it appears on websites, but the sound effect I'm not used to. I like the spinning UI because it appears exactly when I finish and disappears when they respond, so it's easy to tell that the world is generating message}". As for participants who chose auditory cues, they believed it was better as it confirmed the agent heard them, was less distracting, and helped reduce silence awkwardness when the agent was thinking. Some example quotes: (1) "\textit{the spinning ui feels off, although it lets the user know the agent is thinking, it feels a bit inhuman}", (2) "\textit{Because it let me know the response was loading through audio}". Two participants also mentioned thinking that a system error has occurred when they saw the visual wait indicator element of the \textit{Artificial} filler for the first time. 

The split in participants' opinions on the relative importance of auditory and visual filler modalities (\autoref{sec:modality_importance}) suggests that no one specific set of features will work for every user, despite prior work indicating that the combination of gestures and voice utterances is best on average~\cite{kum_can_2022}. We recommend that researchers allow participants to choose their preferred filler modality as long as it is not a factor in their study, and that practitioners allow users to personalize conversational fillers for IVAs. While \textit{Artificial} fillers effectively signaled that the system was processing input, future studies should explore more communicative and socially expressive indicators, such as familiar icons that users associate with thinking, or sequences that represent distinct stages of the response generation process.

\subsection{Implications for Research and Design}
\label{sec:implications}

Based on our findings, we distilled recommendations applicable to future research and design of embodied conversational agents.

\textit{\textbf{Minimize response latency.}}
Minimizing the turn-taking delay in agents' responses is crucial for sustainably-high QoE. Response delays of over 4 seconds worsen participants' perception on broader metrics about embodied agents, skewing collected data and limiting its generalizability.

\textit{\textbf{Choose fillers consistent with the system's purpose.}}
While \textit{Natural} conversational fillers improve perceived system response time, their appropriateness depends on the system's purpose. If accuracy is more important than response speed or being liked by users, \textit{Artificial} fillers or no fillers could fit better.

\textit{\textbf{Allow response filler personalization.}}
Presenting different conversational filler options and allowing users to select among them will account for preferences in visual and auditory modalities. This extends to the granularity of individual animations and phrases, as cultural backgrounds influence the interpretation of gestures. 

\textit{\textbf{Maintain participant engagement.}}
Keeping participants engaged for the entire duration of a user study is important for collecting quality data~\cite{yu_engaging_2020}. Designing quest-like scenarios and minimizing response latency are possible ways to maintain high participant engagement in experiments.

\textit{\textbf{Select a medium that minimizes distractions.}} Immersive VR environments can reduce external visual and auditory distractions~\cite{kuvar_novel_2024}, making them well-suited for studies involving embodied conversational agents. In our experiment, this helped ensure that participant responses were influenced by agent behavior rather than uncontrolled environmental factors.

\section{Limitations and Future Work}

While the questions that we used in our experiment served us to get insight into user perception of IVAs, we acknowledge the need for a more standardized survey to collect perception metrics in this context. Our results serve as an initial comparison point with future work on IVAs. This is apparent in post-condition questions, where all aspects of discomfort and competence metrics from RoSAS~\cite{carpinella_robotic_2017} were aggregated into Q4 and Q5, respectively. This may have reduced the probability of detecting significant differences between conditions. Post-study questions PSQ4, PSQ5 and PSQ10 gauged future use intent; however, the order in which they were presented and the absence of clear application scenarios or reference points may have inadvertently influenced participants' responses. We recommend future work to consider such nuances when designing questionnaires.

The \textit{Natural} and \textit{Artificial} fillers used in our study were context-neutral and appropriate for the chosen scenarios. However, in serious scenarios (e.g., medical), a "laugh" filler would be less appropriate than "let me check your record", and the opposite applies in a playful scenario where human-likeness is more valued. 
In conversations between humans, prosodic features~\cite{heldner_pauses_2010}, conversational fillers, body gestures, and facial expressions, appear before an interlocutor finishes responding~\cite{matsumoto_microexpressions_2018, domaneschi_facial_2017, russell_faces_1997}. The challenge of adding these features to IVAs is rooted in the same reason that IVA response delays exists --- it takes time to predict the appropriate voice line, gesture, and facial expression. Prior work integrated contextual animations for virtual agents, however, they were only triggered after multiple responses of the same type were generated, thus being played too late to mitigate response delay~\cite{zhu_free_form_2023}. Future work should address this by designing fast NLP-based models that would process conversation history within a time shorter than 300ms~\cite{stivers_universals_2009}, or potentially even before the user finishes speaking. 

Applying conversational fillers to other applications of embodied conversational agents could yield valuable insights. For instance, conversational IVAs could \textit{interrupt} humans, mimicking the natural flow of human-human dialogue~\cite{heldner_pauses_2010, stivers_universals_2009, levinson_timing_2015}. As LLMs gain stronger multilingual capabilities~\cite{vayani2024languagesmatterevaluatinglmms}, fillers could be adapted for culturally-sensitive timing in language learning scenarios~\cite{topsakal_framework_2022, divekar_foreign_2022}, potentially increasing IVA acceptance. Time-sensitive settings may also amplify the effects of latency: urgency can compress users' tolerance for delay~\cite{niculescu2010socializing, yang2015effect} and increase cognitive load~\cite{wu2022time}. While our study avoided time pressure by allowing participants to proceed at their own pace, future work should evaluate fillers in high-stakes or time-constrained contexts, such as a desert survival task~\cite{lafferty1974desert}, where embodied assistants have been shown to ease cognitive load~\cite{de2020reducing}. Lastly, we find that research on conversational agents is increasingly conducted in immersive VR, however, there is a need for a more scrutinized evaluation of virtual mediums that can be used for studying ECAs.

\section{Conclusion}
We explored conversational fillers' impact on mitigating response delays in free-form conversations with LLM-powered embodied conversational agents in VR. We found that \textit{Natural} fillers enhance VR user experience by significantly improving participants' perceived response time and reducing latency's negative repercussions. Our findings also indicate that \textit{Artificial} fillers, namely wait indicators and processing sound effects, were not effective at reducing perceived response time. Our results contribute to the nascent work in optimizing user experiences with LLM-powered virtual agents, where network latency or hardware constraints inflate conversational system response time, especially in VR simulations of HRI and HAI scenarios. We outline design recommendations based on our findings and contribute an open-source pipeline as a solution to deploy LLM-based intelligent virtual agents in VR, advancing research efforts in developing more immersive and human-like interactions in VR.

\begin{acks}
We thank the anonymous reviewers for their insightful feedback and the ISUE lab members for their support. This work was in part supported by NSF Award CNS-2326134, DEVCOM Award W912CG2320004, and the Florida High Tech Corridor Council Industry Matching Research Program.
\end{acks}

\bibliographystyle{ACM-Reference-Format}
\bibliography{main}


\begin{thebibliography}{108}


\ifx \showCODEN    \undefined \def \showCODEN     #1{\unskip}     \fi
\ifx \showDOI      \undefined \def \showDOI       #1{#1}\fi
\ifx \showISBNx    \undefined \def \showISBNx     #1{\unskip}     \fi
\ifx \showISBNxiii \undefined \def \showISBNxiii  #1{\unskip}     \fi
\ifx \showISSN     \undefined \def \showISSN      #1{\unskip}     \fi
\ifx \showLCCN     \undefined \def \showLCCN      #1{\unskip}     \fi
\ifx \shownote     \undefined \def \shownote      #1{#1}          \fi
\ifx \showarticletitle \undefined \def \showarticletitle #1{#1}   \fi
\ifx \showURL      \undefined \def \showURL       {\relax}        \fi
\providecommand\bibfield[2]{#2}
\providecommand\bibinfo[2]{#2}
\providecommand\natexlab[1]{#1}
\providecommand\showeprint[2][]{arXiv:#2}

\bibitem[Ali et~al\mbox{.}(2020)]%
        {ali_virtual_2020}
\bibfield{author}{\bibinfo{person}{Mohammad~Rafayet Ali}, \bibinfo{person}{Seyedeh~Zahra Razavi}, \bibinfo{person}{Raina Langevin}, \bibinfo{person}{Abdullah Al~Mamun}, \bibinfo{person}{Benjamin Kane}, \bibinfo{person}{Reza Rawassizadeh}, \bibinfo{person}{Lenhart~K. Schubert}, {and} \bibinfo{person}{Ehsan Hoque}.} \bibinfo{year}{2020}\natexlab{}.
\newblock \showarticletitle{A {Virtual} {Conversational} {Agent} for {Teens} with {Autism} {Spectrum} {Disorder}: {Experimental} {Results} and {Design} {Lessons}}. In \bibinfo{booktitle}{\emph{Proceedings of the 20th {ACM} {International} {Conference} on {Intelligent} {Virtual} {Agents}}} \emph{(\bibinfo{series}{{IVA} '20})}. \bibinfo{publisher}{Association for Computing Machinery}, \bibinfo{address}{New York, NY, USA}, \bibinfo{pages}{1--8}.
\newblock
\showISBNx{978-1-4503-7586-3}
\urldef\tempurl%
\url{https://doi.org/10.1145/3383652.3423900}
\showDOI{\tempurl}


\bibitem[Amer and Johnson(2016)]%
        {amer2016information}
\bibfield{author}{\bibinfo{person}{T.~S. Amer} {and} \bibinfo{person}{Todd~L. Johnson}.} \bibinfo{year}{2016}\natexlab{}.
\newblock \showarticletitle{Information Technology Progress Indicators: Temporal Expectancy, User Preference, and the Perception of Process Duration}.
\newblock \bibinfo{journal}{\emph{International Journal of Technology and Human Interaction (IJTHI)}} \bibinfo{volume}{12}, \bibinfo{number}{4} (\bibinfo{date}{Oct.} \bibinfo{year}{2016}), \bibinfo{pages}{1–--14}.
\newblock
\showISSN{1548-3908}
\urldef\tempurl%
\url{https://doi.org/10.4018/IJTHI.2016100101}
\showDOI{\tempurl}


\bibitem[Aneja et~al\mbox{.}(2021)]%
        {aneja_understanding_2021}
\bibfield{author}{\bibinfo{person}{Deepali Aneja}, \bibinfo{person}{Rens Hoegen}, \bibinfo{person}{Daniel McDuff}, {and} \bibinfo{person}{Mary Czerwinski}.} \bibinfo{year}{2021}\natexlab{}.
\newblock \showarticletitle{Understanding {Conversational} and {Expressive} {Style} in a {Multimodal} {Embodied} {Conversational} {Agent}}. In \bibinfo{booktitle}{\emph{Proceedings of the 2021 {CHI} {Conference} on {Human} {Factors} in {Computing} {Systems}}} \emph{(\bibinfo{series}{{CHI} '21})}. \bibinfo{publisher}{Association for Computing Machinery}, \bibinfo{address}{New York, NY, USA}, \bibinfo{pages}{1--10}.
\newblock
\showISBNx{978-1-4503-8096-6}
\urldef\tempurl%
\url{https://doi.org/10.1145/3411764.3445708}
\showDOI{\tempurl}


\bibitem[Appel et~al\mbox{.}(2012)]%
        {appel2012does}
\bibfield{author}{\bibinfo{person}{Jana Appel}, \bibinfo{person}{Astrid von~der Pütten}, \bibinfo{person}{Nicole~C. Krämer}, {and} \bibinfo{person}{Jonathan Gratch}.} \bibinfo{year}{2012}\natexlab{}.
\newblock \showarticletitle{Does Humanity Matter? Analyzing the Importance of Social Cues and Perceived Agency of a Computer System for the Emergence of Social Reactions during Human-Computer Interaction}.
\newblock \bibinfo{journal}{\emph{Advances in Human-Computer Interaction}} \bibinfo{volume}{2012}, \bibinfo{number}{1} (\bibinfo{year}{2012}), \bibinfo{pages}{324694}.
\newblock
\urldef\tempurl%
\url{https://doi.org/10.1155/2012/324694}
\showDOI{\tempurl}


\bibitem[Bailey et~al\mbox{.}(2019)]%
        {bailey_virtual_2019}
\bibfield{author}{\bibinfo{person}{Jakki~O. Bailey}, \bibinfo{person}{Jeremy~N. Bailenson}, \bibinfo{person}{Jelena Obradović}, {and} \bibinfo{person}{Naomi~R. Aguiar}.} \bibinfo{year}{2019}\natexlab{}.
\newblock \showarticletitle{Virtual Reality's Effect on Children's Inhibitory Control, Social Compliance, and Sharing}.
\newblock \bibinfo{journal}{\emph{Journal of Applied Developmental Psychology}}  \bibinfo{volume}{64} (\bibinfo{date}{July} \bibinfo{year}{2019}), \bibinfo{pages}{101052}.
\newblock
\showISSN{0193-3973}
\urldef\tempurl%
\url{https://doi.org/10.1016/j.appdev.2019.101052}
\showDOI{\tempurl}


\bibitem[Bavelas and Chovil(1997)]%
        {russell_faces_1997}
\bibfield{author}{\bibinfo{person}{Janet~Beavin Bavelas} {and} \bibinfo{person}{Nicole Chovil}.} \bibinfo{year}{1997}\natexlab{}.
\newblock \showarticletitle{Faces in Dialogue}.
\newblock In \bibinfo{booktitle}{\emph{The {Psychology} of {Facial} {Expression}}}, \bibfield{editor}{\bibinfo{person}{James~A. Russell} {and} \bibinfo{person}{José~Miguel Fernández-Dols}} (Eds.). \bibinfo{publisher}{Cambridge University Press}, \bibinfo{address}{Cambridge}, \bibinfo{pages}{334--346}.
\newblock
\showISBNx{978-0-521-58796-9}
\urldef\tempurl%
\url{https://doi.org/10.1017/CBO9780511659911.017}
\showDOI{\tempurl}


\bibitem[Bayat et~al\mbox{.}(2024)]%
        {bayat_exploring_2024}
\bibfield{author}{\bibinfo{person}{Rojin Bayat}, \bibinfo{person}{Elios De~Maio}, \bibinfo{person}{Jacopo Fiorenza}, \bibinfo{person}{Massimo Migliorini}, {and} \bibinfo{person}{Fabrizio Lamberti}.} \bibinfo{year}{2024}\natexlab{}.
\newblock \showarticletitle{Exploring {Methodologies} to {Create} a {Unified} {VR} {User}-{Experience} in the {Field} of {Virtual} {Museum} {Experiences}}. In \bibinfo{booktitle}{\emph{2024 {IEEE} {Gaming}, {Entertainment}, and {Media} {Conference} ({GEM})}}. \bibinfo{publisher}{IEEE}, \bibinfo{address}{Turin, Italy}, \bibinfo{pages}{1--4}.
\newblock
\urldef\tempurl%
\url{https://doi.org/10.1109/GEM61861.2024.10585452}
\showDOI{\tempurl}
\newblock
\shownote{ISSN: 2766-6530}.


\bibitem[Baylor and Rosenberg-Kima(2006)]%
        {baylor2006interface}
\bibfield{author}{\bibinfo{person}{Amy~L. Baylor} {and} \bibinfo{person}{Rinat~B. Rosenberg-Kima}.} \bibinfo{year}{2006}\natexlab{}.
\newblock \showarticletitle{Interface agents to alleviate online frustration}, In \bibinfo{booktitle}{Proceedings of the 7th International Conference on Learning Sciences} (Bloomington, Indiana).
\newblock \bibinfo{journal}{\emph{Proceedings of ICLS 2006}}  \bibinfo{volume}{1}, \bibinfo{pages}{30–36}.
\newblock
\showISBNx{0805861742}
\urldef\tempurl%
\url{https://repository.isls.org//handle/1/3514}
\showURL{%
\tempurl}


\bibitem[Beňuš and Trnka(2014)]%
        {bevnuvs2014prosody}
\bibfield{author}{\bibinfo{person}{Štefan Beňuš} {and} \bibinfo{person}{Marián Trnka}.} \bibinfo{year}{2014}\natexlab{}.
\newblock \showarticletitle{Prosody, Voice Assimilation, and Conversational Fillers}. In \bibinfo{booktitle}{\emph{Speech {Prosody} 2014}}. \bibinfo{publisher}{ISCA}, \bibinfo{address}{Online}, \bibinfo{pages}{75--79}.
\newblock
\showISSN{2333-2042}
\urldef\tempurl%
\url{https://doi.org/10.21437/SpeechProsody.2014-3}
\showDOI{\tempurl}


\bibitem[Bingham(2023)]%
        {bingham_data_2023}
\bibfield{author}{\bibinfo{person}{Andrea~J. Bingham}.} \bibinfo{year}{2023}\natexlab{}.
\newblock \showarticletitle{From {Data} {Management} to {Actionable} {Findings}: {A} {Five}-{Phase} {Process} of {Qualitative} {Data} {Analysis}}.
\newblock \bibinfo{journal}{\emph{International Journal of Qualitative Methods}}  \bibinfo{volume}{22} (\bibinfo{date}{Jan.} \bibinfo{year}{2023}), \bibinfo{pages}{1--11}.
\newblock
\showISSN{1609-4069}
\urldef\tempurl%
\url{https://doi.org/10.1177/16094069231183620}
\showDOI{\tempurl}
\newblock
\shownote{Publisher: SAGE Publications Inc}.


\bibitem[Bortfeld et~al\mbox{.}(2001)]%
        {bortfeld2001disfluency}
\bibfield{author}{\bibinfo{person}{Heather Bortfeld}, \bibinfo{person}{Silvia~D Leon}, \bibinfo{person}{Jonathan~E Bloom}, \bibinfo{person}{Michael~F Schober}, {and} \bibinfo{person}{Susan~E Brennan}.} \bibinfo{year}{2001}\natexlab{}.
\newblock \showarticletitle{Disfluency Rates in Conversation: Effects of Age, Relationship, Topic, Role, and Gender}.
\newblock \bibinfo{journal}{\emph{Language and Speech}} \bibinfo{volume}{44}, \bibinfo{number}{2} (\bibinfo{year}{2001}), \bibinfo{pages}{123--147}.
\newblock
\urldef\tempurl%
\url{https://doi.org/10.1177/00238309010440020101}
\showDOI{\tempurl}
\newblock
\shownote{PMID: 11575901}.


\bibitem[Boukaram et~al\mbox{.}(2021)]%
        {boukaram2021mitigating}
\bibfield{author}{\bibinfo{person}{Halim-Antoine Boukaram}, \bibinfo{person}{Micheline Ziadee}, {and} \bibinfo{person}{Majd~F Sakr}.} \bibinfo{year}{2021}\natexlab{}.
\newblock \showarticletitle{Mitigating the {Effects} of {Delayed} {Virtual} {Agent} {Response} {Time} {Using} {Conversational} {Fillers}}. In \bibinfo{booktitle}{\emph{Proceedings of the 9th {International} {Conference} on {Human}-{Agent} {Interaction}}} \emph{(\bibinfo{series}{{HAI} '21})}. \bibinfo{publisher}{Association for Computing Machinery}, \bibinfo{address}{New York, NY, USA}, \bibinfo{pages}{130--138}.
\newblock
\showISBNx{978-1-4503-8620-3}
\urldef\tempurl%
\url{https://doi.org/10.1145/3472307.3484181}
\showDOI{\tempurl}


\bibitem[Branaghan and Sanchez(2009)]%
        {branaghan2009feedback}
\bibfield{author}{\bibinfo{person}{Russell~J Branaghan} {and} \bibinfo{person}{Christopher~A Sanchez}.} \bibinfo{year}{2009}\natexlab{}.
\newblock \showarticletitle{Feedback Preferences and Impressions of Waiting}.
\newblock \bibinfo{journal}{\emph{Human Factors}} \bibinfo{volume}{51}, \bibinfo{number}{4} (\bibinfo{year}{2009}), \bibinfo{pages}{528--538}.
\newblock
\urldef\tempurl%
\url{https://doi.org/10.1177/0018720809345684}
\showDOI{\tempurl}
\newblock
\shownote{PMID: 19899362}.


\bibitem[Carpinella et~al\mbox{.}(2017)]%
        {carpinella_robotic_2017}
\bibfield{author}{\bibinfo{person}{Colleen~M. Carpinella}, \bibinfo{person}{Alisa~B. Wyman}, \bibinfo{person}{Michael~A. Perez}, {and} \bibinfo{person}{Steven~J. Stroessner}.} \bibinfo{year}{2017}\natexlab{}.
\newblock \showarticletitle{The {Robotic} {Social} {Attributes} {Scale} ({RoSAS}): {Development} and {Validation}}. In \bibinfo{booktitle}{\emph{Proceedings of the 2017 {ACM}/{IEEE} {International} {Conference} on {Human}-{Robot} {Interaction}}} \emph{(\bibinfo{series}{{HRI} '17})}. \bibinfo{publisher}{Association for Computing Machinery}, \bibinfo{address}{New York, NY, USA}, \bibinfo{pages}{254--262}.
\newblock
\showISBNx{978-1-4503-4336-7}
\urldef\tempurl%
\url{https://doi.org/10.1145/2909824.3020208}
\showDOI{\tempurl}


\bibitem[Casas et~al\mbox{.}(2024)]%
        {casas_moodflow_2024}
\bibfield{author}{\bibinfo{person}{Llogari Casas}, \bibinfo{person}{Samantha Hannah}, {and} \bibinfo{person}{Kenny Mitchell}.} \bibinfo{year}{2024}\natexlab{}.
\newblock \showarticletitle{{MoodFlow}: {Orchestrating} {Conversations} with {Emotionally} {Intelligent} {Avatars} in {Mixed} {Reality}}. In \bibinfo{booktitle}{\emph{2024 {IEEE} {Conference} on {Virtual} {Reality} and {3D} {User} {Interfaces} {Abstracts} and {Workshops} ({VRW})}}. \bibinfo{publisher}{IEEE}, \bibinfo{address}{Orlando, FL, USA}, \bibinfo{pages}{86--89}.
\newblock
\urldef\tempurl%
\url{https://doi.org/10.1109/VRW62533.2024.00021}
\showDOI{\tempurl}


\bibitem[Charlier et~al\mbox{.}(2022)]%
        {florian_charlier_2022_7213391}
\bibfield{author}{\bibinfo{person}{Florian Charlier}, \bibinfo{person}{Marc Weber}, \bibinfo{person}{Dariusz Izak}, \bibinfo{person}{Emerson Harkin}, \bibinfo{person}{Marcin Magnus}, \bibinfo{person}{Joseph Lalli}, \bibinfo{person}{Louison Fresnais}, \bibinfo{person}{Matt Chan}, \bibinfo{person}{Nikolay Markov}, \bibinfo{person}{Oren Amsalem}, \bibinfo{person}{Sebastian Proost}, \bibinfo{person}{Agamemnon Krasoulis}, \bibinfo{person}{getzze}, {and} \bibinfo{person}{Stefan Repplinger}.} \bibinfo{year}{2022}\natexlab{}.
\newblock \bibinfo{booktitle}{\emph{Statannotations}}.
\newblock Statannotations Contributors.
\newblock
\urldef\tempurl%
\url{https://doi.org/10.5281/zenodo.7213391}
\showDOI{\tempurl}


\bibitem[Cheng et~al\mbox{.}(2024)]%
        {cheng2024effects}
\bibfield{author}{\bibinfo{person}{Anping Cheng}, \bibinfo{person}{Dongming Ma}, \bibinfo{person}{Hao Qian}, {and} \bibinfo{person}{Younghwan Pan}.} \bibinfo{year}{2024}\natexlab{}.
\newblock \showarticletitle{The Effects of Mobile Applications’ Passive and Interactive Loading Screen Types on Waiting Experience}.
\newblock \bibinfo{journal}{\emph{Behaviour \& Information Technology}} \bibinfo{volume}{43}, \bibinfo{number}{8} (\bibinfo{year}{2024}), \bibinfo{pages}{1652--1663}.
\newblock
\urldef\tempurl%
\url{https://doi.org/10.1080/0144929X.2023.2224901}
\showDOI{\tempurl}


\bibitem[Chheang et~al\mbox{.}(2024)]%
        {chheang_towards_2024}
\bibfield{author}{\bibinfo{person}{Vuthea Chheang}, \bibinfo{person}{Shayla Sharmin}, \bibinfo{person}{Rommy Márquez-Hernández}, \bibinfo{person}{Megha Patel}, \bibinfo{person}{Danush Rajasekaran}, \bibinfo{person}{Gavin Caulfield}, \bibinfo{person}{Behdokht Kiafar}, \bibinfo{person}{Jicheng Li}, \bibinfo{person}{Pinar Kullu}, {and} \bibinfo{person}{Roghayeh~Leila Barmaki}.} \bibinfo{year}{2024}\natexlab{}.
\newblock \showarticletitle{Towards {Anatomy} {Education} with {Generative} {AI}-based {Virtual} {Assistants} in {Immersive} {Virtual} {Reality} {Environments}}. In \bibinfo{booktitle}{\emph{2024 {IEEE} {International} {Conference} on {Artificial} {Intelligence} and {eXtended} and {Virtual} {Reality} ({AIxVR})}}. \bibinfo{publisher}{IEEE}, \bibinfo{address}{Piscataway, NJ, USA}, \bibinfo{pages}{21--30}.
\newblock
\urldef\tempurl%
\url{https://doi.org/10.1109/AIxVR59861.2024.00011}
\showDOI{\tempurl}
\newblock
\shownote{ISSN: 2771-7453}.


\bibitem[Christenfeld(1995)]%
        {christenfeld1995does}
\bibfield{author}{\bibinfo{person}{Nicholas Christenfeld}.} \bibinfo{year}{1995}\natexlab{}.
\newblock \showarticletitle{Does It Hurt to Say Um?}
\newblock \bibinfo{journal}{\emph{Journal of Nonverbal Behavior}}  \bibinfo{volume}{19} (\bibinfo{year}{1995}), \bibinfo{pages}{171--186}.
\newblock
\urldef\tempurl%
\url{https://doi.org/10.1007/BF02175503}
\showDOI{\tempurl}


\bibitem[Clark(1996)]%
        {clark1996using}
\bibfield{author}{\bibinfo{person}{Herbert~H Clark}.} \bibinfo{year}{1996}\natexlab{}.
\newblock \bibinfo{booktitle}{\emph{Using language}}.
\newblock \bibinfo{publisher}{Cambridge University Press}, \bibinfo{address}{Cambridge, UK}.
\newblock
\showISBNx{9780511620539}
\urldef\tempurl%
\url{https://doi.org/10.1017/CBO9780511620539}
\showDOI{\tempurl}


\bibitem[Conrad et~al\mbox{.}(2010)]%
        {conrad2010impact}
\bibfield{author}{\bibinfo{person}{Frederick~G Conrad}, \bibinfo{person}{Mick~P Couper}, \bibinfo{person}{Roger Tourangeau}, {and} \bibinfo{person}{Andy Peytchev}.} \bibinfo{year}{2010}\natexlab{}.
\newblock \showarticletitle{The Impact of Progress Indicators on Task Completion}.
\newblock \bibinfo{journal}{\emph{Interacting with computers}} \bibinfo{volume}{22}, \bibinfo{number}{5} (\bibinfo{year}{2010}), \bibinfo{pages}{417--427}.
\newblock
\urldef\tempurl%
\url{https://doi.org/10.1016/j.intcom.2010.03.001}
\showDOI{\tempurl}


\bibitem[Creem-Regehr et~al\mbox{.}(2022)]%
        {creem_regehr_perceiving_2022}
\bibfield{author}{\bibinfo{person}{Sarah~H. Creem-Regehr}, \bibinfo{person}{Jeanine~K. Stefanucci}, {and} \bibinfo{person}{Bobby Bodenheimer}.} \bibinfo{year}{2022}\natexlab{}.
\newblock \showarticletitle{Perceiving Distance in Virtual Reality: Theoretical Insights from Contemporary Technologies}.
\newblock \bibinfo{journal}{\emph{Philosophical Transactions of the Royal Society B: Biological Sciences}} \bibinfo{volume}{378}, \bibinfo{number}{1869} (\bibinfo{date}{Dec.} \bibinfo{year}{2022}), \bibinfo{pages}{1--12}.
\newblock
\urldef\tempurl%
\url{https://doi.org/10.1098/rstb.2021.0456}
\showDOI{\tempurl}


\bibitem[de~Melo et~al\mbox{.}(2020)]%
        {de2020reducing}
\bibfield{author}{\bibinfo{person}{Celso~M. de Melo}, \bibinfo{person}{Kangsoo Kim}, \bibinfo{person}{Nahal Norouzi}, \bibinfo{person}{Gerd Bruder}, {and} \bibinfo{person}{Gregory Welch}.} \bibinfo{year}{2020}\natexlab{}.
\newblock \showarticletitle{Reducing {Cognitive} {Load} and {Improving} {Warfighter} {Problem} {Solving} {With} {Intelligent} {Virtual} {Assistants}}.
\newblock \bibinfo{journal}{\emph{Frontiers in Psychology}}  \bibinfo{volume}{11} (\bibinfo{year}{2020}), \bibinfo{numpages}{12}~pages.
\newblock
\showISSN{1664-1078}
\urldef\tempurl%
\url{https://doi.org/10.3389/fpsyg.2020.554706}
\showDOI{\tempurl}


\bibitem[Divekar* et~al\mbox{.}(2022)]%
        {divekar_foreign_2022}
\bibfield{author}{\bibinfo{person}{Rahul~R. Divekar*}, \bibinfo{person}{Jaimie Drozdal*}, \bibinfo{person}{Samuel Chabot*}, \bibinfo{person}{Yalun Zhou}, \bibinfo{person}{Hui Su}, \bibinfo{person}{Yue Chen}, \bibinfo{person}{Houming Zhu}, \bibinfo{person}{James~A. Hendler}, {and} \bibinfo{person}{Jonas Braasch}.} \bibinfo{year}{2022}\natexlab{}.
\newblock \showarticletitle{Foreign Language Acquisition via Artificial Intelligence and Extended Reality: Design and Evaluation}.
\newblock \bibinfo{journal}{\emph{Computer Assisted Language Learning}} \bibinfo{volume}{35}, \bibinfo{number}{9} (\bibinfo{date}{Dec} \bibinfo{year}{2022}), \bibinfo{pages}{2332--2360}.
\newblock
\showISSN{0958-8221}
\urldef\tempurl%
\url{https://doi.org/10.1080/09588221.2021.1879162}
\showDOI{\tempurl}


\bibitem[Do et~al\mbox{.}(2023)]%
        {doValid2023}
\bibfield{author}{\bibinfo{person}{Tiffany~D. Do}, \bibinfo{person}{Steve Zelenty}, \bibinfo{person}{Mar Gonzalez-Franco}, {and} \bibinfo{person}{Ryan~P. McMahan}.} \bibinfo{year}{2023}\natexlab{}.
\newblock \showarticletitle{{VALID}: A Perceptually Validated {Virtual} {Avatar} {Library} for {Inclusion} and {Diversity}}.
\newblock \bibinfo{journal}{\emph{Frontiers in Virtual Reality}}  \bibinfo{volume}{4} (\bibinfo{date}{Nov.} \bibinfo{year}{2023}), \bibinfo{numpages}{15}~pages.
\newblock
\showISSN{2673-4192}
\urldef\tempurl%
\url{https://doi.org/10.3389/frvir.2023.1248915}
\showDOI{\tempurl}


\bibitem[Domaneschi et~al\mbox{.}(2017)]%
        {domaneschi_facial_2017}
\bibfield{author}{\bibinfo{person}{Filippo Domaneschi}, \bibinfo{person}{Marcello Passarelli}, {and} \bibinfo{person}{Carlo Chiorri}.} \bibinfo{year}{2017}\natexlab{}.
\newblock \showarticletitle{Facial Expressions and Speech Acts: Experimental Evidences on the Role of the Upper Face as an Illocutionary Force Indicating Device in Language Comprehension}.
\newblock \bibinfo{journal}{\emph{Cognitive Processing}} \bibinfo{volume}{18}, \bibinfo{number}{3} (\bibinfo{date}{Aug.} \bibinfo{year}{2017}), \bibinfo{pages}{285--306}.
\newblock
\showISSN{1612-4790}
\urldef\tempurl%
\url{https://doi.org/10.1007/s10339-017-0809-6}
\showDOI{\tempurl}


\bibitem[Elkin et~al\mbox{.}(2021)]%
        {artc}
\bibfield{author}{\bibinfo{person}{Lisa~A. Elkin}, \bibinfo{person}{Matthew Kay}, \bibinfo{person}{James~J. Higgins}, {and} \bibinfo{person}{Jacob~O. Wobbrock}.} \bibinfo{year}{2021}\natexlab{}.
\newblock \showarticletitle{An Aligned Rank Transform Procedure for Multifactor Contrast Tests}. In \bibinfo{booktitle}{\emph{The 34th Annual ACM Symposium on User Interface Software and Technology}} (Virtual Event, USA) \emph{(\bibinfo{series}{UIST '21})}. \bibinfo{publisher}{Association for Computing Machinery}, \bibinfo{address}{New York, NY, USA}, \bibinfo{pages}{754–768}.
\newblock
\showISBNx{9781450386357}
\urldef\tempurl%
\url{https://doi.org/10.1145/3472749.3474784}
\showDOI{\tempurl}


\bibitem[Faul et~al\mbox{.}(2009)]%
        {faul2009statistical}
\bibfield{author}{\bibinfo{person}{Franz Faul}, \bibinfo{person}{Edgar Erdfelder}, \bibinfo{person}{Axel Buchner}, {and} \bibinfo{person}{Albert-Georg Lang}.} \bibinfo{year}{2009}\natexlab{}.
\newblock \showarticletitle{Statistical Power Analyses Using {G}*{Power} 3.1: {Tests} for Correlation and Regression Analyses}.
\newblock \bibinfo{journal}{\emph{Behavior Research Methods}} \bibinfo{volume}{41}, \bibinfo{number}{4} (\bibinfo{date}{Nov} \bibinfo{year}{2009}), \bibinfo{pages}{1149--1160}.
\newblock
\showISSN{1554-3528}
\urldef\tempurl%
\url{https://doi.org/10.3758/BRM.41.4.1149}
\showDOI{\tempurl}


\bibitem[Funakoshi et~al\mbox{.}(2008)]%
        {funakoshi_smoothing_2008}
\bibfield{author}{\bibinfo{person}{Kotaro Funakoshi}, \bibinfo{person}{Kazuki Kobayashi}, \bibinfo{person}{Mikio Nakano}, \bibinfo{person}{Seiji Yamada}, \bibinfo{person}{Yasuhiko Kitamura}, {and} \bibinfo{person}{Hiroshi Tsujino}.} \bibinfo{year}{2008}\natexlab{}.
\newblock \showarticletitle{Smoothing Human-Robot Speech Interactions by Using a Blinking-light as Subtle Expression}. In \bibinfo{booktitle}{\emph{Proceedings of the 10th international conference on {Multimodal} interfaces}} \emph{(\bibinfo{series}{{ICMI} '08})}. \bibinfo{publisher}{Association for Computing Machinery}, \bibinfo{address}{New York, NY, USA}, \bibinfo{pages}{293--296}.
\newblock
\showISBNx{978-1-60558-198-9}
\urldef\tempurl%
\url{https://doi.org/10.1145/1452392.1452452}
\showDOI{\tempurl}


\bibitem[Funk et~al\mbox{.}(2020)]%
        {funk2020usable}
\bibfield{author}{\bibinfo{person}{Markus Funk}, \bibinfo{person}{Carie Cunningham}, \bibinfo{person}{Duygu Kanver}, \bibinfo{person}{Christopher Saikalis}, {and} \bibinfo{person}{Rohan Pansare}.} \bibinfo{year}{2020}\natexlab{}.
\newblock \showarticletitle{Usable and {Acceptable} {Response} {Delays} of {Conversational} {Agents} in {Automotive} {User} {Interfaces}}. In \bibinfo{booktitle}{\emph{12th {International} {Conference} on {Automotive} {User} {Interfaces} and {Interactive} {Vehicular} {Applications}}} \emph{(\bibinfo{series}{{AutomotiveUI} '20})}. \bibinfo{publisher}{Association for Computing Machinery}, \bibinfo{address}{New York, NY, USA}, \bibinfo{pages}{262--269}.
\newblock
\showISBNx{978-1-4503-8065-2}
\urldef\tempurl%
\url{https://doi.org/10.1145/3409120.3410651}
\showDOI{\tempurl}


\bibitem[Garcia et~al\mbox{.}(2024)]%
        {garcia_speaking_2024}
\bibfield{author}{\bibinfo{person}{Irene~Lopez Garcia}, \bibinfo{person}{Ephraim Schott}, \bibinfo{person}{Marcel Gohsen}, \bibinfo{person}{Volker Bernhard}, \bibinfo{person}{Benno Stein}, {and} \bibinfo{person}{Bernd Froehlich}.} \bibinfo{year}{2024}\natexlab{}.
\newblock \showarticletitle{Speaking with {Objects}: {Conversational} {Agents}’ {Embodiment} in {Virtual} {Museums}}. In \bibinfo{booktitle}{\emph{2024 {IEEE} {International} {Symposium} on {Mixed} and {Augmented} {Reality} ({ISMAR})}}. \bibinfo{publisher}{IEEE}, \bibinfo{address}{Piscataway, NJ, USA}, \bibinfo{pages}{279--288}.
\newblock
\urldef\tempurl%
\url{https://doi.org/10.1109/ISMAR62088.2024.00042}
\showDOI{\tempurl}
\newblock
\shownote{ISSN: 2473-0726}.


\bibitem[Gnewuch et~al\mbox{.}(2018a)]%
        {gnewuch2018faster}
\bibfield{author}{\bibinfo{person}{Ulrich Gnewuch}, \bibinfo{person}{Stefan Morana}, \bibinfo{person}{Marc Adam}, {and} \bibinfo{person}{Alexander Maedche}.} \bibinfo{year}{2018}\natexlab{a}.
\newblock \showarticletitle{Faster is Not Always Better: Understanding the Effect of Dynamic Response Delays in Human-Chatbot Interaction}. In \bibinfo{booktitle}{\emph{European Conference on Information Systems}}. \bibinfo{publisher}{AIS Electronic Library (AISeL)}, \bibinfo{address}{Portsmouth, UK}, \bibinfo{pages}{1--17}.
\newblock
\urldef\tempurl%
\url{https://aisel.aisnet.org/ecis2018_rp/113}
\showURL{%
\tempurl}


\bibitem[Gnewuch et~al\mbox{.}(2018b)]%
        {gnewuch2018chatbot}
\bibfield{author}{\bibinfo{person}{Ulrich Gnewuch}, \bibinfo{person}{Stefan Morana}, \bibinfo{person}{Marc Adam}, {and} \bibinfo{person}{Alexander Maedche}.} \bibinfo{year}{2018}\natexlab{b}.
\newblock \showarticletitle{“{The} {Chatbot} is typing ...” – {The} {Role} of {Typing} {Indicators} in {Human}-{Chatbot} {Interaction}}. In \bibinfo{booktitle}{\emph{SIGHCI 2018 Proceedings}}. \bibinfo{publisher}{AIS Electronic Library (AISeL)}, \bibinfo{address}{San Francisco, CA}, \bibinfo{pages}{1--5}.
\newblock
\urldef\tempurl%
\url{https://aisel.aisnet.org/sighci2018/14}
\showURL{%
\tempurl}


\bibitem[Gonzalez-Franco et~al\mbox{.}(2020)]%
        {rocketbox}
\bibfield{author}{\bibinfo{person}{Mar Gonzalez-Franco}, \bibinfo{person}{Eyal Ofek}, \bibinfo{person}{Ye Pan}, \bibinfo{person}{Angus Antley}, \bibinfo{person}{Anthony Steed}, \bibinfo{person}{Bernhard Spanlang}, \bibinfo{person}{Antonella Maselli}, \bibinfo{person}{Domna Banakou}, \bibinfo{person}{Nuria Pelechano}, \bibinfo{person}{Sergio Orts-Escolano}, \bibinfo{person}{Veronica Orvalho}, \bibinfo{person}{Laura Trutoiu}, \bibinfo{person}{Markus Wojcik}, \bibinfo{person}{Maria~V. Sanchez-Vives}, \bibinfo{person}{Jeremy Bailenson}, \bibinfo{person}{Mel Slater}, {and} \bibinfo{person}{Jaron Lanier}.} \bibinfo{year}{2020}\natexlab{}.
\newblock \showarticletitle{The Rocketbox Library and the Utility of Freely Available Rigged Avatars}.
\newblock \bibinfo{journal}{\emph{Frontiers in Virtual Reality}}  \bibinfo{volume}{1} (\bibinfo{year}{2020}), \bibinfo{numpages}{20}~pages.
\newblock
\showISSN{2673-4192}
\urldef\tempurl%
\url{https://doi.org/10.3389/frvir.2020.561558}
\showDOI{\tempurl}


\bibitem[Harrison et~al\mbox{.}(2007)]%
        {harrison2007rethinking}
\bibfield{author}{\bibinfo{person}{Chris Harrison}, \bibinfo{person}{Brian Amento}, \bibinfo{person}{Stacey Kuznetsov}, {and} \bibinfo{person}{Robert Bell}.} \bibinfo{year}{2007}\natexlab{}.
\newblock \showarticletitle{Rethinking the Progress Bar}. In \bibinfo{booktitle}{\emph{Proceedings of the 20th annual {ACM} symposium on {User} interface software and technology}} \emph{(\bibinfo{series}{{UIST} '07})}. \bibinfo{publisher}{Association for Computing Machinery}, \bibinfo{address}{New York, NY, USA}, \bibinfo{pages}{115--118}.
\newblock
\showISBNx{978-1-59593-679-0}
\urldef\tempurl%
\url{https://doi.org/10.1145/1294211.1294231}
\showDOI{\tempurl}


\bibitem[Harrison et~al\mbox{.}(2010)]%
        {harrison2010faster}
\bibfield{author}{\bibinfo{person}{Chris Harrison}, \bibinfo{person}{Zhiquan Yeo}, {and} \bibinfo{person}{Scott~E. Hudson}.} \bibinfo{year}{2010}\natexlab{}.
\newblock \showarticletitle{Faster Progress Bars: Manipulating Perceived Duration with Visual Augmentations}. In \bibinfo{booktitle}{\emph{Proceedings of the {SIGCHI} {Conference} on {Human} {Factors} in {Computing} {Systems}}} \emph{(\bibinfo{series}{{CHI} '10})}. \bibinfo{publisher}{Association for Computing Machinery}, \bibinfo{address}{New York, NY, USA}, \bibinfo{pages}{1545--1548}.
\newblock
\showISBNx{978-1-60558-929-9}
\urldef\tempurl%
\url{https://doi.org/10.1145/1753326.1753556}
\showDOI{\tempurl}


\bibitem[Hasan et~al\mbox{.}(2023)]%
        {hasan_sapien_2023}
\bibfield{author}{\bibinfo{person}{Masum Hasan}, \bibinfo{person}{Cengiz Ozel}, \bibinfo{person}{Sammy Potter}, {and} \bibinfo{person}{Ehsan Hoque}.} \bibinfo{year}{2023}\natexlab{}.
\newblock \showarticletitle{{SAPIEN}: {Affective} {Virtual} {Agents} {Powered} by {Large} {Language} {Models}*}. In \bibinfo{booktitle}{\emph{2023 11th {International} {Conference} on {Affective} {Computing} and {Intelligent} {Interaction} {Workshops} and {Demos} ({ACIIW})}}. \bibinfo{publisher}{IEEE}, \bibinfo{address}{Cambridge, MA, USA}, \bibinfo{pages}{1--3}.
\newblock
\urldef\tempurl%
\url{https://doi.org/10.1109/ACIIW59127.2023.10388188}
\showDOI{\tempurl}


\bibitem[Heldner and Edlund(2010)]%
        {heldner_pauses_2010}
\bibfield{author}{\bibinfo{person}{Mattias Heldner} {and} \bibinfo{person}{Jens Edlund}.} \bibinfo{year}{2010}\natexlab{}.
\newblock \showarticletitle{Pauses, Gaps and Overlaps in Conversations}.
\newblock \bibinfo{journal}{\emph{Journal of Phonetics}} \bibinfo{volume}{38}, \bibinfo{number}{4} (\bibinfo{date}{Oct.} \bibinfo{year}{2010}), \bibinfo{pages}{555--568}.
\newblock
\showISSN{0095-4470}
\urldef\tempurl%
\url{https://doi.org/10.1016/j.wocn.2010.08.002}
\showDOI{\tempurl}


\bibitem[Hohenstein et~al\mbox{.}(2016)]%
        {hohenstein2016shorter}
\bibfield{author}{\bibinfo{person}{Jess Hohenstein}, \bibinfo{person}{Hani Khan}, \bibinfo{person}{Kramer Canfield}, \bibinfo{person}{Samuel Tung}, {and} \bibinfo{person}{Rocio Perez~Cano}.} \bibinfo{year}{2016}\natexlab{}.
\newblock \showarticletitle{Shorter {Wait} {Times}: {The} {Effects} of {Various} {Loading} {Screens} on {Perceived} {Performance}}. In \bibinfo{booktitle}{\emph{Proceedings of the 2016 {CHI} {Conference} {Extended} {Abstracts} on {Human} {Factors} in {Computing} {Systems}}} \emph{(\bibinfo{series}{{CHI} {EA} '16})}. \bibinfo{publisher}{Association for Computing Machinery}, \bibinfo{address}{New York, NY, USA}, \bibinfo{pages}{3084--3090}.
\newblock
\showISBNx{978-1-4503-4082-3}
\urldef\tempurl%
\url{https://doi.org/10.1145/2851581.2892308}
\showDOI{\tempurl}


\bibitem[Holtgraves et~al\mbox{.}(2007)]%
        {holtgraves_perceiving_2007}
\bibfield{author}{\bibinfo{person}{T.~M. Holtgraves}, \bibinfo{person}{S.~J. Ross}, \bibinfo{person}{C.~R. Weywadt}, {and} \bibinfo{person}{T.~L. Han}.} \bibinfo{year}{2007}\natexlab{}.
\newblock \showarticletitle{Perceiving Artificial Social Agents}.
\newblock \bibinfo{journal}{\emph{Computers in Human Behavior}} \bibinfo{volume}{23}, \bibinfo{number}{5} (\bibinfo{date}{Sept.} \bibinfo{year}{2007}), \bibinfo{pages}{2163--2174}.
\newblock
\showISSN{0747-5632}
\urldef\tempurl%
\url{https://doi.org/10.1016/j.chb.2006.02.017}
\showDOI{\tempurl}


\bibitem[Hone(2006)]%
        {hone2006empathic}
\bibfield{author}{\bibinfo{person}{Kate Hone}.} \bibinfo{year}{2006}\natexlab{}.
\newblock \showarticletitle{Empathic Agents to Reduce User Frustration: The Effects of Varying Agent Characteristics}.
\newblock \bibinfo{journal}{\emph{Interacting with computers}} \bibinfo{volume}{18}, \bibinfo{number}{2} (\bibinfo{year}{2006}), \bibinfo{pages}{227--245}.
\newblock
\urldef\tempurl%
\url{https://doi.org/10.1016/j.intcom.2005.05.003}
\showDOI{\tempurl}


\bibitem[Hornik(1984)]%
        {hornik1984subjective}
\bibfield{author}{\bibinfo{person}{Jacob Hornik}.} \bibinfo{year}{1984}\natexlab{}.
\newblock \showarticletitle{Subjective vs. Objective Time Measures: A Note on the Perception of Time in Consumer Behavior}.
\newblock \bibinfo{journal}{\emph{Journal of consumer research}} \bibinfo{volume}{11}, \bibinfo{number}{1} (\bibinfo{year}{1984}), \bibinfo{pages}{615--618}.
\newblock
\urldef\tempurl%
\url{https://doi.org/10.1086/208998}
\showDOI{\tempurl}


\bibitem[Hoxmeier and DiCesare(2000)]%
        {hoxmeier2000system}
\bibfield{author}{\bibinfo{person}{John Hoxmeier} {and} \bibinfo{person}{Chris DiCesare}.} \bibinfo{year}{2000}\natexlab{}.
\newblock \bibinfo{title}{System {Response} {Time} and {User} {Satisfaction}: {An} {Experimental} {Study} of {Browser}-based {Applications}}.
\newblock , \bibinfo{numpages}{6}~pages.
\newblock
\urldef\tempurl%
\url{https://aisel.aisnet.org/amcis2000/347}
\showURL{%
\tempurl}


\bibitem[Hu and Pan(2024)]%
        {hu_is_2024}
\bibfield{author}{\bibinfo{person}{Qian Hu} {and} \bibinfo{person}{Zhao Pan}.} \bibinfo{year}{2024}\natexlab{}.
\newblock \showarticletitle{Is Cute {AI} More Forgivable? {The} Impact of Informal Language Styles and Relationship Norms of Conversational Agents on Service Recovery}.
\newblock \bibinfo{journal}{\emph{Electronic Commerce Research and Applications}}  \bibinfo{volume}{65} (\bibinfo{date}{May} \bibinfo{year}{2024}), \bibinfo{pages}{1--14}.
\newblock
\showISSN{1567-4223}
\urldef\tempurl%
\url{https://doi.org/10.1016/j.elerap.2024.101398}
\showDOI{\tempurl}


\bibitem[Huang and Lo(2025)]%
        {huang_human_2025}
\bibfield{author}{\bibinfo{person}{Zuwen Huang} {and} \bibinfo{person}{Ada Lo}.} \bibinfo{year}{2025}\natexlab{}.
\newblock \showarticletitle{Human vs. Robot Service Provider Agents in Service Failures: Comparing Customer Dissatisfaction and The Mediating Role of Forgiveness and Service Recovery Expectation}.
\newblock \bibinfo{journal}{\emph{Information Technology \& Tourism}}  \bibinfo{volume}{27} (\bibinfo{date}{Feb.} \bibinfo{year}{2025}), \bibinfo{pages}{1--32}.
\newblock
\showISSN{1943-4294}
\urldef\tempurl%
\url{https://doi.org/10.1007/s40558-025-00314-6}
\showDOI{\tempurl}


\bibitem[Hwang et~al\mbox{.}(2019)]%
        {hwang_when_2019}
\bibfield{author}{\bibinfo{person}{Sun~Young Hwang}, \bibinfo{person}{Negar Khojasteh}, {and} \bibinfo{person}{Susan~R. Fussell}.} \bibinfo{year}{2019}\natexlab{}.
\newblock \showarticletitle{When {Delayed} in a {Hurry}: {Interpretations} of {Response} {Delays} in {Time}-{Sensitive} {Instant} {Messaging}}.
\newblock \bibinfo{journal}{\emph{Proceedings of the ACM on Human-Computer Interaction}} \bibinfo{volume}{3}, \bibinfo{number}{GROUP} (\bibinfo{date}{Dec.} \bibinfo{year}{2019}), \bibinfo{pages}{1--20}.
\newblock
\showISSN{2573-0142}
\urldef\tempurl%
\url{https://doi.org/10.1145/3361115}
\showDOI{\tempurl}


\bibitem[Iftikhar et~al\mbox{.}(2023)]%
        {iftikhar2023together}
\bibfield{author}{\bibinfo{person}{Zainab Iftikhar}, \bibinfo{person}{Yumeng Ma}, {and} \bibinfo{person}{Jeff Huang}.} \bibinfo{year}{2023}\natexlab{}.
\newblock \showarticletitle{“Together But Not Together”: Evaluating Typing Indicators for Interaction-Rich Communication}. In \bibinfo{booktitle}{\emph{Proceedings of the 2023 CHI Conference on Human Factors in Computing Systems}} (Hamburg, Germany) \emph{(\bibinfo{series}{CHI '23})}. \bibinfo{publisher}{Association for Computing Machinery}, \bibinfo{address}{New York, NY, USA}, Article \bibinfo{articleno}{724}, \bibinfo{numpages}{12}~pages.
\newblock
\showISBNx{9781450394215}
\urldef\tempurl%
\url{https://doi.org/10.1145/3544548.3581248}
\showDOI{\tempurl}


\bibitem[Jaffe and Feldstein(1970)]%
        {jaffe1970rhythms}
\bibfield{author}{\bibinfo{person}{Joseph Jaffe} {and} \bibinfo{person}{Stanley Feldstein}.} \bibinfo{year}{1970}\natexlab{}.
\newblock \bibinfo{booktitle}{\emph{Rhythms of Dialogue}}.
\newblock \bibinfo{publisher}{Academic Press}, \bibinfo{address}{New York, NY, USA}.
\newblock
\showISBNx{0123798507}
\urldef\tempurl%
\url{https://cir.nii.ac.jp/crid/1130282269381241984}
\showURL{%
\tempurl}


\bibitem[Jeong et~al\mbox{.}(2019)]%
        {jeong2019exploring}
\bibfield{author}{\bibinfo{person}{Yuin Jeong}, \bibinfo{person}{Juho Lee}, {and} \bibinfo{person}{Younah Kang}.} \bibinfo{year}{2019}\natexlab{}.
\newblock \showarticletitle{Exploring {Effects} of {Conversational} {Fillers} on {User} {Perception} of {Conversational} {Agents}}. In \bibinfo{booktitle}{\emph{Extended {Abstracts} of the 2019 {CHI} {Conference} on {Human} {Factors} in {Computing} {Systems}}} \emph{(\bibinfo{series}{{CHI} {EA} '19})}. \bibinfo{publisher}{Association for Computing Machinery}, \bibinfo{address}{New York, NY, USA}, \bibinfo{numpages}{6}~pages.
\newblock
\showISBNx{978-1-4503-5971-9}
\urldef\tempurl%
\url{https://doi.org/10.1145/3290607.3312913}
\showDOI{\tempurl}


\bibitem[Kanda et~al\mbox{.}(2007)]%
        {kanda_humanoid_2007}
\bibfield{author}{\bibinfo{person}{Takayuki Kanda}, \bibinfo{person}{Masayuki Kamasima}, \bibinfo{person}{Michita Imai}, \bibinfo{person}{Tetsuo Ono}, \bibinfo{person}{Daisuke Sakamoto}, \bibinfo{person}{Hiroshi Ishiguro}, {and} \bibinfo{person}{Yuichiro Anzai}.} \bibinfo{year}{2007}\natexlab{}.
\newblock \showarticletitle{A Humanoid Robot that Pretends to Listen to Route Guidance from a Human}.
\newblock \bibinfo{journal}{\emph{Auton. Robots}} \bibinfo{volume}{22}, \bibinfo{number}{1} (\bibinfo{date}{Jan.} \bibinfo{year}{2007}), \bibinfo{pages}{87--100}.
\newblock
\showISSN{0929-5593}
\urldef\tempurl%
\url{https://doi.org/10.1007/s10514-006-9007-6}
\showDOI{\tempurl}


\bibitem[Kim et~al\mbox{.}(2017)]%
        {kim2017effect}
\bibfield{author}{\bibinfo{person}{Woojoo Kim}, \bibinfo{person}{Shuping Xiong}, {and} \bibinfo{person}{Zhuoqian Liang}.} \bibinfo{year}{2017}\natexlab{}.
\newblock \showarticletitle{Effect of Loading Symbol of Online Video on Perception of Waiting Time}.
\newblock \bibinfo{journal}{\emph{International Journal of Human--Computer Interaction}} \bibinfo{volume}{33}, \bibinfo{number}{12} (\bibinfo{year}{2017}), \bibinfo{pages}{1001--1009}.
\newblock
\urldef\tempurl%
\url{https://doi.org/10.1080/10447318.2017.1305051}
\showDOI{\tempurl}


\bibitem[Komatsu et~al\mbox{.}(2024)]%
        {komatsu2024waiting}
\bibfield{author}{\bibinfo{person}{Takanori Komatsu}, \bibinfo{person}{Chenxi Xie}, {and} \bibinfo{person}{Seiji Yamada}.} \bibinfo{year}{2024}\natexlab{}.
\newblock \showarticletitle{Waiting {Time} {Perceptions} for {Faster} {Count}-downs/ups {Are} {More} {Sensitive} {Than} {Slower} {Ones}: {Experimental} {Investigation} and {Its} {Application}}. In \bibinfo{booktitle}{\emph{Proceedings of the {CHI} {Conference} on {Human} {Factors} in {Computing} {Systems}}} \emph{(\bibinfo{series}{{CHI} '24})}. \bibinfo{publisher}{Association for Computing Machinery}, \bibinfo{address}{New York, NY, USA}, \bibinfo{pages}{1--13}.
\newblock
\showISBNx{9798400703300}
\urldef\tempurl%
\url{https://doi.org/10.1145/3613904.3641942}
\showDOI{\tempurl}


\bibitem[Kum and Lee(2022)]%
        {kum_can_2022}
\bibfield{author}{\bibinfo{person}{Junyeong Kum} {and} \bibinfo{person}{Myungho Lee}.} \bibinfo{year}{2022}\natexlab{}.
\newblock \showarticletitle{Can {Gestural} {Filler} {Reduce} {User}-{Perceived} {Latency} in {Conversation} with {Digital} {Humans}?}
\newblock \bibinfo{journal}{\emph{Applied Sciences}} \bibinfo{volume}{12}, \bibinfo{number}{21} (\bibinfo{date}{Jan.} \bibinfo{year}{2022}), \bibinfo{pages}{10972}.
\newblock
\showISSN{2076-3417}
\urldef\tempurl%
\url{https://doi.org/10.3390/app122110972}
\showDOI{\tempurl}


\bibitem[Kuvar et~al\mbox{.}(2024)]%
        {kuvar_novel_2024}
\bibfield{author}{\bibinfo{person}{Vishal~Kiran Kuvar}, \bibinfo{person}{Jeremy~N. Bailenson}, {and} \bibinfo{person}{Caitlin Mills}.} \bibinfo{year}{2024}\natexlab{}.
\newblock \showarticletitle{A novel quantitative assessment of engagement in virtual reality: {Task}-unrelated thought is reduced compared to {2D} videos.}
\newblock \bibinfo{journal}{\emph{Computers \& Education}}  \bibinfo{volume}{209} (\bibinfo{date}{Feb.} \bibinfo{year}{2024}), \bibinfo{pages}{104959}.
\newblock
\showISSN{0360-1315}
\urldef\tempurl%
\url{https://doi.org/10.1016/j.compedu.2023.104959}
\showDOI{\tempurl}


\bibitem[Kyrlitsias and Michael-Grigoriou(2022)]%
        {kyrlitsias_social_2022}
\bibfield{author}{\bibinfo{person}{Christos Kyrlitsias} {and} \bibinfo{person}{Despina Michael-Grigoriou}.} \bibinfo{year}{2022}\natexlab{}.
\newblock \showarticletitle{Social {Interaction} {With} {Agents} and {Avatars} in {Immersive} {Virtual} {Environments}: {A} {Survey}}.
\newblock \bibinfo{journal}{\emph{Frontiers in Virtual Reality}}  \bibinfo{volume}{2} (\bibinfo{date}{Jan.} \bibinfo{year}{2022}), \bibinfo{numpages}{13}~pages.
\newblock
\showISSN{2673-4192}
\urldef\tempurl%
\url{https://doi.org/10.3389/frvir.2021.786665}
\showDOI{\tempurl}
\newblock
\shownote{Publisher: Frontiers}.


\bibitem[Lafferty et~al\mbox{.}(1974)]%
        {lafferty1974desert}
\bibfield{author}{\bibinfo{person}{J.C. Lafferty}, \bibinfo{person}{P.M. Eady}, \bibinfo{person}{A.W. Pond}, {and} \bibinfo{person}{Human Synergistics}.} \bibinfo{year}{1974}\natexlab{}.
\newblock \bibinfo{booktitle}{\emph{The Desert Survival Problem: A Group Decision Making Experience for Examining and Increasing Individual and Team Effectiveness: Manual}}.
\newblock \bibinfo{publisher}{Experimental Learning Methods}, \bibinfo{address}{Plymouth, Michigan: Experimental Learning Methods}.
\newblock
\urldef\tempurl%
\url{https://books.google.com/books?id=X37-GwAACAAJ}
\showURL{%
\tempurl}


\bibitem[Laghari et~al\mbox{.}(2019)]%
        {laghari_application_2019}
\bibfield{author}{\bibinfo{person}{Asif~Ali Laghari}, \bibinfo{person}{Hui He}, \bibinfo{person}{Muhammad Shafiq}, {and} \bibinfo{person}{Asiya Khan}.} \bibinfo{year}{2019}\natexlab{}.
\newblock \showarticletitle{Application of {Quality} of {Experience} in {Networked} {Services}: {Review}, {Trend} \& {Perspectives}}.
\newblock \bibinfo{journal}{\emph{Systemic Practice and Action Research}} \bibinfo{volume}{32}, \bibinfo{number}{5} (\bibinfo{date}{Oct.} \bibinfo{year}{2019}), \bibinfo{pages}{501--519}.
\newblock
\showISSN{1573-9295}
\urldef\tempurl%
\url{https://doi.org/10.1007/s11213-018-9471-x}
\showDOI{\tempurl}


\bibitem[Levinson and Torreira(2015)]%
        {levinson_timing_2015}
\bibfield{author}{\bibinfo{person}{Stephen~C. Levinson} {and} \bibinfo{person}{Francisco Torreira}.} \bibinfo{year}{2015}\natexlab{}.
\newblock \showarticletitle{Timing in Turn-taking and its Implications for Processing Models of Language}.
\newblock \bibinfo{journal}{\emph{Frontiers in Psychology}}  \bibinfo{volume}{6} (\bibinfo{year}{2015}), \bibinfo{pages}{731}.
\newblock
\showISSN{1664-1078}
\urldef\tempurl%
\url{https://doi.org/10.3389/fpsyg.2015.00731}
\showDOI{\tempurl}


\bibitem[Lew et~al\mbox{.}(2018)]%
        {lew_interactivity_2018}
\bibfield{author}{\bibinfo{person}{Zijian Lew}, \bibinfo{person}{Joseph~B Walther}, \bibinfo{person}{Augustine Pang}, {and} \bibinfo{person}{Wonsun Shin}.} \bibinfo{year}{2018}\natexlab{}.
\newblock \showarticletitle{Interactivity in {Online} {Chat}: {Conversational} {Contingency} and {Response} {Latency} in {Computer}-mediated {Communication}}.
\newblock \bibinfo{journal}{\emph{Journal of Computer-Mediated Communication}} \bibinfo{volume}{23}, \bibinfo{number}{4} (\bibinfo{date}{July} \bibinfo{year}{2018}), \bibinfo{pages}{201--221}.
\newblock
\showISSN{1083-6101}
\urldef\tempurl%
\url{https://doi.org/10.1093/jcmc/zmy009}
\showDOI{\tempurl}


\bibitem[Li and Chen(2019a)]%
        {li2019effect}
\bibfield{author}{\bibinfo{person}{Shasha Li} {and} \bibinfo{person}{Chien-Hsiung Chen}.} \bibinfo{year}{2019}\natexlab{a}.
\newblock \showarticletitle{The {Effect} of {Progress} {Indicator} {Speeds} on {Users}’ {Time} {Perceptions} and {Experience} of a {Smartphone} {User} {Interface}}. In \bibinfo{booktitle}{\emph{Human-{Computer} {Interaction}. {Recognition} and {Interaction} {Technologies}: {Thematic} {Area}, {HCI} 2019, {Held} as {Part} of the 21st {HCI} {International} {Conference}, {HCII} 2019, {Orlando}, {FL}, {USA}, {July} 26–31, 2019, {Proceedings}, {Part} {II}}}. \bibinfo{publisher}{Springer-Verlag}, \bibinfo{address}{Berlin, Heidelberg}, \bibinfo{pages}{28--36}.
\newblock
\showISBNx{978-3-030-22642-8}
\urldef\tempurl%
\url{https://doi.org/10.1007/978-3-030-22643-5_3}
\showDOI{\tempurl}


\bibitem[Li and Chen(2019b)]%
        {li2019effects}
\bibfield{author}{\bibinfo{person}{Shasha Li} {and} \bibinfo{person}{Chien-Hsiung Chen}.} \bibinfo{year}{2019}\natexlab{b}.
\newblock \showarticletitle{The Effects of Visual Feedback Designs on Long Wait Time of Mobile Application User Interface}.
\newblock \bibinfo{journal}{\emph{Interacting with Computers}} \bibinfo{volume}{31}, \bibinfo{number}{1} (\bibinfo{date}{Jan.} \bibinfo{year}{2019}), \bibinfo{pages}{1--12}.
\newblock
\showISSN{0953-5438}
\urldef\tempurl%
\url{https://doi.org/10.1093/iwc/iwz001}
\showDOI{\tempurl}


\bibitem[Llanes-Jurado et~al\mbox{.}(2024)]%
        {llanes_jurado_developing_2024}
\bibfield{author}{\bibinfo{person}{Jose Llanes-Jurado}, \bibinfo{person}{Lucía Gómez-Zaragozá}, \bibinfo{person}{Maria~Eleonora Minissi}, \bibinfo{person}{Mariano Alcañiz}, {and} \bibinfo{person}{Javier Marín-Morales}.} \bibinfo{year}{2024}\natexlab{}.
\newblock \showarticletitle{Developing Conversational {Virtual} {Humans} for Social Emotion Elicitation Based on Large Language Models}.
\newblock \bibinfo{journal}{\emph{Expert Systems with Applications}}  \bibinfo{volume}{246} (\bibinfo{date}{Jul} \bibinfo{year}{2024}), \bibinfo{pages}{123261}.
\newblock
\showISSN{0957-4174}
\urldef\tempurl%
\url{https://doi.org/10.1016/j.eswa.2024.123261}
\showDOI{\tempurl}


\bibitem[L{\'o}pez~Gambino et~al\mbox{.}(2017)]%
        {gambino2017beyond}
\bibfield{author}{\bibinfo{person}{Soledad L{\'o}pez~Gambino}, \bibinfo{person}{Sina Zarrie{\ss}}, {and} \bibinfo{person}{David Schlangen}.} \bibinfo{year}{2017}\natexlab{}.
\newblock \showarticletitle{Beyond On-hold Messages: Conversational Time-buying in Task-oriented Dialogue}. In \bibinfo{booktitle}{\emph{Proceedings of the 18th Annual {SIG}dial Meeting on Discourse and Dialogue}}, \bibfield{editor}{\bibinfo{person}{Kristiina Jokinen}, \bibinfo{person}{Manfred Stede}, \bibinfo{person}{David DeVault}, {and} \bibinfo{person}{Annie Louis}} (Eds.). \bibinfo{publisher}{Association for Computational Linguistics}, \bibinfo{address}{Saarbr{\"u}cken, Germany}, \bibinfo{pages}{241--246}.
\newblock
\urldef\tempurl%
\url{https://doi.org/10.18653/v1/W17-5529}
\showDOI{\tempurl}


\bibitem[López~Gambino et~al\mbox{.}(2019)]%
        {lopez2019testing}
\bibfield{author}{\bibinfo{person}{Soledad López~Gambino}, \bibinfo{person}{Sina Zarrieß}, {and} \bibinfo{person}{David Schlangen}.} \bibinfo{year}{2019}\natexlab{}.
\newblock \showarticletitle{Testing {Strategies} {For} {Bridging} {Time}-{To}-{Content} {In} {Spoken} {Dialogue} {Systems}}. In \bibinfo{booktitle}{\emph{9th {International} {Workshop} on {Spoken} {Dialogue} {System} {Technology}}}, \bibfield{editor}{\bibinfo{person}{Luis~Fernando D'Haro}, \bibinfo{person}{Rafael~E. Banchs}, {and} \bibinfo{person}{Haizhou Li}} (Eds.). Springer, \bibinfo{publisher}{Springer}, \bibinfo{address}{Singapore}, \bibinfo{pages}{103--109}.
\newblock
\showISBNx{9789811394430}
\urldef\tempurl%
\url{https://doi.org/10.1007/978-981-13-9443-0_9}
\showDOI{\tempurl}


\bibitem[Mahmud et~al\mbox{.}(2019)]%
        {mahmud_quality_2019}
\bibfield{author}{\bibinfo{person}{Redowan Mahmud}, \bibinfo{person}{Satish~Narayana Srirama}, \bibinfo{person}{Kotagiri Ramamohanarao}, {and} \bibinfo{person}{Rajkumar Buyya}.} \bibinfo{year}{2019}\natexlab{}.
\newblock \showarticletitle{Quality of {Experience} ({QoE})-aware Placement of Applications in {Fog} Computing Environments}.
\newblock \bibinfo{journal}{\emph{J. Parallel and Distrib. Comput.}}  \bibinfo{volume}{132} (\bibinfo{date}{Oct.} \bibinfo{year}{2019}), \bibinfo{pages}{190--203}.
\newblock
\showISSN{0743-7315}
\urldef\tempurl%
\url{https://doi.org/10.1016/j.jpdc.2018.03.004}
\showDOI{\tempurl}


\bibitem[Maslych et~al\mbox{.}(2024a)]%
        {maslych2024takeaways}
\bibfield{author}{\bibinfo{person}{Mykola Maslych}, \bibinfo{person}{Christian Pumarada}, \bibinfo{person}{Amirpouya Ghasemaghaei}, {and} \bibinfo{person}{Joseph J.~LaViola Jr}.} \bibinfo{year}{2024}\natexlab{a}.
\newblock \bibinfo{title}{Takeaways from Applying LLM Capabilities to Multiple Conversational Avatars in a VR Pilot Study}.
\newblock
\newblock
\showeprint[arxiv]{2501.00168}~[cs.HC]


\bibitem[Maslych et~al\mbox{.}(2024b)]%
        {maslych2024selectionsdatabase}
\bibfield{author}{\bibinfo{person}{Mykola Maslych}, \bibinfo{person}{Difeng Yu}, \bibinfo{person}{Amirpouya Ghasemaghaei}, \bibinfo{person}{Yahya Hmaiti}, \bibinfo{person}{Esteban~Segarra Martinez}, \bibinfo{person}{Dominic Simon}, \bibinfo{person}{Eugene~Matthew Taranta}, \bibinfo{person}{Joanna Bergstr{\"o}m}, {and} \bibinfo{person}{Joseph~J. LaViola~Jr}.} \bibinfo{year}{2024}\natexlab{b}.
\newblock \showarticletitle{From Research to Practice: Survey and Taxonomy of Object Selection in Consumer VR Applications}. In \bibinfo{booktitle}{\emph{2024 IEEE International Symposium on Mixed and Augmented Reality (ISMAR)}} (Seattle, WA, USA) \emph{(\bibinfo{series}{ISMAR '24})}. \bibinfo{publisher}{IEEE}, \bibinfo{address}{Piscataway, NJ, USA}, \bibinfo{pages}{990--999}.
\newblock
\urldef\tempurl%
\url{https://doi.org/10.1109/ISMAR62088.2024.00115}
\showDOI{\tempurl}


\bibitem[Matsumoto and Hwang(2018)]%
        {matsumoto_microexpressions_2018}
\bibfield{author}{\bibinfo{person}{David Matsumoto} {and} \bibinfo{person}{Hyisung~C. Hwang}.} \bibinfo{year}{2018}\natexlab{}.
\newblock \showarticletitle{Microexpressions {Differentiate} {Truths} {From} {Lies} {About} {Future} {Malicious} {Intent}}.
\newblock \bibinfo{journal}{\emph{Frontiers in Psychology}}  \bibinfo{volume}{9} (\bibinfo{date}{Dec.} \bibinfo{year}{2018}), \bibinfo{pages}{2545}.
\newblock
\showISSN{1664-1078}
\urldef\tempurl%
\url{https://doi.org/10.3389/fpsyg.2018.02545}
\showDOI{\tempurl}


\bibitem[McWilliams et~al\mbox{.}(2015)]%
        {mcwilliams_secondary_2015}
\bibfield{author}{\bibinfo{person}{Thomas McWilliams}, \bibinfo{person}{Bryan Reimer}, \bibinfo{person}{Bruce Mehler}, \bibinfo{person}{Jonathan Dobres}, {and} \bibinfo{person}{Hale McAnulty}.} \bibinfo{year}{2015}\natexlab{}.
\newblock \showarticletitle{A {Secondary} {Assessment} of the {Impact} of {Voice} {Interface} {Turn} {Delays} on {Driver} {Attention} and {Arousal} in {Field} {Conditions}}.
\newblock \bibinfo{journal}{\emph{Driving Assessment Conference}} \bibinfo{volume}{8}, \bibinfo{number}{2015} (\bibinfo{date}{June} \bibinfo{year}{2015}), \bibinfo{pages}{408--414}.
\newblock
\urldef\tempurl%
\url{https://pubs.lib.uiowa.edu/driving/article/id/28561/}
\showURL{%
\tempurl}


\bibitem[Miller(1968)]%
        {miller_response_1968}
\bibfield{author}{\bibinfo{person}{Robert~B. Miller}.} \bibinfo{year}{1968}\natexlab{}.
\newblock \showarticletitle{Response Time in Man-computer Conversational Transactions}. In \bibinfo{booktitle}{\emph{Proceedings of the {December} 9-11, 1968, fall joint computer conference, part {I} on - {AFIPS} '68 ({Fall}, part {I})}}. \bibinfo{publisher}{ACM Press}, \bibinfo{address}{San Francisco, California}, \bibinfo{pages}{267}.
\newblock
\urldef\tempurl%
\url{https://doi.org/10.1145/1476589.1476628}
\showDOI{\tempurl}


\bibitem[Myers(1985)]%
        {myers1985importance}
\bibfield{author}{\bibinfo{person}{Brad~A Myers}.} \bibinfo{year}{1985}\natexlab{}.
\newblock \showarticletitle{The Importance of Percent-Done Progress Indicators for Computer-Human Interfaces}.
\newblock \bibinfo{journal}{\emph{ACM SIGCHI Bulletin}} \bibinfo{volume}{16}, \bibinfo{number}{4} (\bibinfo{year}{1985}), \bibinfo{pages}{11--17}.
\newblock
\showISSN{0736-6906}
\urldef\tempurl%
\url{https://doi.org/10.1145/1165385.317459}
\showDOI{\tempurl}


\bibitem[Müller-Brockhausen et~al\mbox{.}(2023)]%
        {muller_brockhausen_chatter_2023}
\bibfield{author}{\bibinfo{person}{Matthias Müller-Brockhausen}, \bibinfo{person}{Giulio Barbero}, {and} \bibinfo{person}{Mike Preuss}.} \bibinfo{year}{2023}\natexlab{}.
\newblock \showarticletitle{Chatter {Generation} through {Language} {Models}}. In \bibinfo{booktitle}{\emph{2023 {IEEE} {Conference} on {Games} ({CoG})}}. \bibinfo{publisher}{IEEE}, \bibinfo{address}{Boston, MA, USA}, \bibinfo{pages}{6}.
\newblock
\urldef\tempurl%
\url{https://doi.org/10.1109/CoG57401.2023.10333244}
\showDOI{\tempurl}
\newblock
\shownote{ISSN: 2325-4289}.


\bibitem[Niculescu et~al\mbox{.}(2010)]%
        {niculescu2010socializing}
\bibfield{author}{\bibinfo{person}{Andreea Niculescu}, \bibinfo{person}{Betsy van Dijk}, \bibinfo{person}{Anton Nijholt}, \bibinfo{person}{Dilip~Kumar Limbu}, \bibinfo{person}{Swee~Lan See}, {and} \bibinfo{person}{Alvin Hong~Yee Wong}.} \bibinfo{year}{2010}\natexlab{}.
\newblock \showarticletitle{Socializing with {Olivia}, the {Youngest} {Robot} {Receptionist} {Outside} the {Lab}}. In \bibinfo{booktitle}{\emph{Social {Robotics}}}, \bibfield{editor}{\bibinfo{person}{Shuzhi~Sam Ge}, \bibinfo{person}{Haizhou Li}, \bibinfo{person}{John-John Cabibihan}, {and} \bibinfo{person}{Yeow~Kee Tan}} (Eds.). Springer, \bibinfo{publisher}{Springer}, \bibinfo{address}{Berlin, Heidelberg}, \bibinfo{pages}{50--62}.
\newblock
\showISBNx{978-3-642-17248-9}
\urldef\tempurl%
\url{https://doi.org/10.1007/978-3-642-17248-9_6}
\showDOI{\tempurl}


\bibitem[Ohshima et~al\mbox{.}(2015)]%
        {ohshima_conversational_2015}
\bibfield{author}{\bibinfo{person}{Naoki Ohshima}, \bibinfo{person}{Keita Kimijima}, \bibinfo{person}{Junji Yamato}, {and} \bibinfo{person}{Naoki Mukawa}.} \bibinfo{year}{2015}\natexlab{}.
\newblock \showarticletitle{A Conversational Robot with Vocal and Bodily Fillers for Recovering from Awkward Silence at Turn-takings}. In \bibinfo{booktitle}{\emph{2015 24th {IEEE} {International} {Symposium} on {Robot} and {Human} {Interactive} {Communication} ({RO}-{MAN})}}. \bibinfo{publisher}{IEEE}, \bibinfo{address}{Kobe, Japan}, \bibinfo{pages}{325--330}.
\newblock
\urldef\tempurl%
\url{https://doi.org/10.1109/ROMAN.2015.7333677}
\showDOI{\tempurl}


\bibitem[Pan and Hamilton(2018)]%
        {pan_why_2018}
\bibfield{author}{\bibinfo{person}{Xueni Pan} {and} \bibinfo{person}{Antonia F. de~C. Hamilton}.} \bibinfo{year}{2018}\natexlab{}.
\newblock \showarticletitle{Why and How to Use Virtual Reality to Study Human Social Interaction: {The} Challenges of Exploring a New Research Landscape}.
\newblock \bibinfo{journal}{\emph{British Journal of Psychology}} \bibinfo{volume}{109}, \bibinfo{number}{3} (\bibinfo{year}{2018}), \bibinfo{pages}{395--417}.
\newblock
\showISSN{2044-8295}
\urldef\tempurl%
\url{https://doi.org/10.1111/bjop.12290}
\showDOI{\tempurl}


\bibitem[Park et~al\mbox{.}(2023)]%
        {park_generative_2023}
\bibfield{author}{\bibinfo{person}{Joon~Sung Park}, \bibinfo{person}{Joseph~C. O'Brien}, \bibinfo{person}{Carrie~J. Cai}, \bibinfo{person}{Meredith~Ringel Morris}, \bibinfo{person}{Percy Liang}, {and} \bibinfo{person}{Michael~S. Bernstein}.} \bibinfo{year}{2023}\natexlab{}.
\newblock \bibinfo{title}{Generative {Agents}: {Interactive} {Simulacra} of {Human} {Behavior}}.
\newblock
\newblock
\showeprint[arxiv]{2304.03442}~[cs.HC]


\bibitem[Petersen et~al\mbox{.}(2021)]%
        {petersen_pedagogical_2021}
\bibfield{author}{\bibinfo{person}{Gustav~Bøg Petersen}, \bibinfo{person}{Aske Mottelson}, {and} \bibinfo{person}{Guido Makransky}.} \bibinfo{year}{2021}\natexlab{}.
\newblock \showarticletitle{Pedagogical {Agents} in {Educational} {VR}: {An} in the {Wild} {Study}}. In \bibinfo{booktitle}{\emph{Proceedings of the 2021 {CHI} {Conference} on {Human} {Factors} in {Computing} {Systems}}} \emph{(\bibinfo{series}{{CHI} '21})}. \bibinfo{publisher}{Association for Computing Machinery}, \bibinfo{address}{New York, NY, USA}, \bibinfo{pages}{1--12}.
\newblock
\showISBNx{978-1-4503-8096-6}
\urldef\tempurl%
\url{https://doi.org/10.1145/3411764.3445760}
\showDOI{\tempurl}


\bibitem[Porcheron et~al\mbox{.}(2018)]%
        {porcheron_voice_2018}
\bibfield{author}{\bibinfo{person}{Martin Porcheron}, \bibinfo{person}{Joel~E. Fischer}, \bibinfo{person}{Stuart Reeves}, {and} \bibinfo{person}{Sarah Sharples}.} \bibinfo{year}{2018}\natexlab{}.
\newblock \showarticletitle{Voice {Interfaces} in {Everyday} {Life}}. In \bibinfo{booktitle}{\emph{Proceedings of the 2018 {CHI} {Conference} on {Human} {Factors} in {Computing} {Systems}}}. \bibinfo{publisher}{ACM}, \bibinfo{address}{Montreal QC Canada}, \bibinfo{pages}{1--12}.
\newblock
\showISBNx{978-1-4503-5620-6}
\urldef\tempurl%
\url{https://doi.org/10.1145/3173574.3174214}
\showDOI{\tempurl}


\bibitem[Qin et~al\mbox{.}(2024)]%
        {qin_charactermeet_2024}
\bibfield{author}{\bibinfo{person}{Hua~Xuan Qin}, \bibinfo{person}{Shan Jin}, \bibinfo{person}{Ze Gao}, \bibinfo{person}{Mingming Fan}, {and} \bibinfo{person}{Pan Hui}.} \bibinfo{year}{2024}\natexlab{}.
\newblock \showarticletitle{{CharacterMeet}: {Supporting} {Creative} {Writers}' {Entire} {Story} {Character} {Construction} {Processes} {Through} {Conversation} with {LLM}-{Powered} {Chatbot} {Avatars}}. In \bibinfo{booktitle}{\emph{Proceedings of the {CHI} {Conference} on {Human} {Factors} in {Computing} {Systems}}} \emph{(\bibinfo{series}{{CHI} '24})}. \bibinfo{publisher}{Association for Computing Machinery}, \bibinfo{address}{New York, NY, USA}, \bibinfo{pages}{1--19}.
\newblock
\showISBNx{9798400703300}
\urldef\tempurl%
\url{https://doi.org/10.1145/3613904.3642105}
\showDOI{\tempurl}


\bibitem[Saeed et~al\mbox{.}(2024)]%
        {saeed_developing_2024}
\bibfield{author}{\bibinfo{person}{Amir~Bani Saeed}, \bibinfo{person}{Zahra Moussavi}, {and} \bibinfo{person}{Bruce Hardy}.} \bibinfo{year}{2024}\natexlab{}.
\newblock \showarticletitle{Developing an {Avatar} in {Virtual} {Reality} for {Mental} {Health} {Treatment}}.
\newblock \bibinfo{journal}{\emph{CMBES Proceedings}}  \bibinfo{volume}{46} (\bibinfo{date}{June} \bibinfo{year}{2024}), \bibinfo{pages}{1--1}.
\newblock
\showISSN{2371-9516}
\urldef\tempurl%
\url{https://proceedings.cmbes.ca/index.php/proceedings/article/view/1107}
\showURL{%
\tempurl}


\bibitem[Shaikh et~al\mbox{.}(2010)]%
        {shaikh_quality_2010}
\bibfield{author}{\bibinfo{person}{Junaid Shaikh}, \bibinfo{person}{Markus Fiedler}, {and} \bibinfo{person}{Denis Collange}.} \bibinfo{year}{2010}\natexlab{}.
\newblock \showarticletitle{Quality of {Experience} from User and Network Perspectives}.
\newblock \bibinfo{journal}{\emph{Annals of Telecommunications - Annales Des Télécommunications}} \bibinfo{volume}{65}, \bibinfo{number}{1} (\bibinfo{date}{Feb.} \bibinfo{year}{2010}), \bibinfo{pages}{47--57}.
\newblock
\showISSN{1958-9395}
\urldef\tempurl%
\url{https://doi.org/10.1007/s12243-009-0142-x}
\showDOI{\tempurl}


\bibitem[Shechtman and Horowitz(2003)]%
        {shechtman2003media}
\bibfield{author}{\bibinfo{person}{Nicole Shechtman} {and} \bibinfo{person}{Leonard~M. Horowitz}.} \bibinfo{year}{2003}\natexlab{}.
\newblock \showarticletitle{Media Inequality in Conversation: How People Behave Differently When Interacting with Computers and People}. In \bibinfo{booktitle}{\emph{Proceedings of the {SIGCHI} {Conference} on {Human} {Factors} in {Computing} {Systems}}} \emph{(\bibinfo{series}{{CHI} '03})}. \bibinfo{publisher}{Association for Computing Machinery}, \bibinfo{address}{New York, NY, USA}, \bibinfo{pages}{281--288}.
\newblock
\showISBNx{978-1-58113-630-2}
\urldef\tempurl%
\url{https://doi.org/10.1145/642611.642661}
\showDOI{\tempurl}


\bibitem[Shiwa et~al\mbox{.}(2009)]%
        {shiwa_how_2009}
\bibfield{author}{\bibinfo{person}{Toshiyuki Shiwa}, \bibinfo{person}{Takayuki Kanda}, \bibinfo{person}{Michita Imai}, \bibinfo{person}{Hiroshi Ishiguro}, {and} \bibinfo{person}{Norihiro Hagita}.} \bibinfo{year}{2009}\natexlab{}.
\newblock \showarticletitle{How {Quickly} {Should} a {Communication} {Robot} {Respond}? {Delaying} {Strategies} and {Habituation} {Effects}}.
\newblock \bibinfo{journal}{\emph{International Journal of Social Robotics}} \bibinfo{volume}{1}, \bibinfo{number}{2} (\bibinfo{date}{April} \bibinfo{year}{2009}), \bibinfo{pages}{141--155}.
\newblock
\showISSN{1875-4805}
\urldef\tempurl%
\url{https://doi.org/10.1007/s12369-009-0012-8}
\showDOI{\tempurl}


\bibitem[Shoa et~al\mbox{.}(2023)]%
        {shoa_sushi_2023}
\bibfield{author}{\bibinfo{person}{Alon Shoa}, \bibinfo{person}{Ramon Oliva}, \bibinfo{person}{Mel Slater}, {and} \bibinfo{person}{Doron Friedman}.} \bibinfo{year}{2023}\natexlab{}.
\newblock \showarticletitle{Sushi with {Einstein}: {Enhancing} {Hybrid} {Live} {Events} with {LLM}-{Based} {Virtual} {Humans}}. In \bibinfo{booktitle}{\emph{Proceedings of the 23rd {ACM} {International} {Conference} on {Intelligent} {Virtual} {Agents}}} \emph{(\bibinfo{series}{{IVA} '23})}. \bibinfo{publisher}{Association for Computing Machinery}, \bibinfo{address}{New York, NY, USA}, \bibinfo{numpages}{6}~pages.
\newblock
\showISBNx{978-1-4503-9994-4}
\urldef\tempurl%
\url{https://doi.org/10.1145/3570945.3607317}
\showDOI{\tempurl}


\bibitem[Skantze and Hjalmarsson(2013)]%
        {skantze2013towards}
\bibfield{author}{\bibinfo{person}{Gabriel Skantze} {and} \bibinfo{person}{Anna Hjalmarsson}.} \bibinfo{year}{2013}\natexlab{}.
\newblock \showarticletitle{Towards Incremental Speech Generation in Conversational Systems}.
\newblock \bibinfo{journal}{\emph{Computer Speech \& Language}} \bibinfo{volume}{27}, \bibinfo{number}{1} (\bibinfo{year}{2013}), \bibinfo{pages}{243--262}.
\newblock
\urldef\tempurl%
\url{https://doi.org/10.1016/j.csl.2012.05.004}
\showDOI{\tempurl}


\bibitem[Srinidhi et~al\mbox{.}(2024)]%
        {srinidhi2024xair}
\bibfield{author}{\bibinfo{person}{Sruti Srinidhi}, \bibinfo{person}{Edward Lu}, {and} \bibinfo{person}{Anthony Rowe}.} \bibinfo{year}{2024}\natexlab{}.
\newblock \showarticletitle{XaiR: An XR Platform that Integrates Large Language Models with the Physical World}. In \bibinfo{booktitle}{\emph{2024 IEEE International Symposium on Mixed and Augmented Reality (ISMAR)}}. \bibinfo{publisher}{IEEE}, \bibinfo{address}{Bellevue, WA, USA}, \bibinfo{pages}{759--767}.
\newblock
\urldef\tempurl%
\url{https://doi.org/10.1109/ISMAR62088.2024.00091}
\showDOI{\tempurl}


\bibitem[Steenstra et~al\mbox{.}(2024)]%
        {steenstra_virtual_2024}
\bibfield{author}{\bibinfo{person}{Ian Steenstra}, \bibinfo{person}{Farnaz Nouraei}, \bibinfo{person}{Mehdi Arjmand}, {and} \bibinfo{person}{Timothy Bickmore}.} \bibinfo{year}{2024}\natexlab{}.
\newblock \showarticletitle{Virtual {Agents} for {Alcohol} {Use} {Counseling}: {Exploring} {LLM}-{Powered} {Motivational} {Interviewing}}. In \bibinfo{booktitle}{\emph{Proceedings of the {ACM} {International} {Conference} on {Intelligent} {Virtual} {Agents}}}. \bibinfo{publisher}{ACM}, \bibinfo{address}{GLASGOW United Kingdom}, \bibinfo{pages}{1--10}.
\newblock
\showISBNx{9798400706257}
\urldef\tempurl%
\url{https://doi.org/10.1145/3652988.3673932}
\showDOI{\tempurl}


\bibitem[Stivers et~al\mbox{.}(2009)]%
        {stivers_universals_2009}
\bibfield{author}{\bibinfo{person}{Tanya Stivers}, \bibinfo{person}{N.~J. Enfield}, \bibinfo{person}{Penelope Brown}, \bibinfo{person}{Christina Englert}, \bibinfo{person}{Makoto Hayashi}, \bibinfo{person}{Trine Heinemann}, \bibinfo{person}{Gertie Hoymann}, \bibinfo{person}{Federico Rossano}, \bibinfo{person}{Jan~Peter de Ruiter}, \bibinfo{person}{Kyung-Eun Yoon}, {and} \bibinfo{person}{Stephen~C. Levinson}.} \bibinfo{year}{2009}\natexlab{}.
\newblock \showarticletitle{Universals and Cultural Variation in Turn-taking in Conversation}.
\newblock \bibinfo{journal}{\emph{Proceedings of the National Academy of Sciences}} \bibinfo{volume}{106}, \bibinfo{number}{26} (\bibinfo{date}{June} \bibinfo{year}{2009}), \bibinfo{pages}{10587--10592}.
\newblock
\urldef\tempurl%
\url{https://doi.org/10.1073/pnas.0903616106}
\showDOI{\tempurl}


\bibitem[Svartvik(1990)]%
        {svartvik1990london}
\bibfield{editor}{\bibinfo{person}{Jan Svartvik}} (Ed.). \bibinfo{year}{1990}\natexlab{}.
\newblock \bibinfo{booktitle}{\emph{The London–Lund Corpus of Spoken English: Description and Research}}. \bibinfo{series}{Lund Studies in English}, Vol.~\bibinfo{volume}{82}.
\newblock \bibinfo{publisher}{Lund University Press}, \bibinfo{address}{Lund, Sweden}.
\newblock
\showISBNx{91-7966-126-2}
\urldef\tempurl%
\url{https://lup.lub.lu.se/record/c9ccd3ca-4a6e-4885-9ca5-1f939baa977f}
\showURL{%
\tempurl}


\bibitem[Swerts(1998)]%
        {swerts1998filled}
\bibfield{author}{\bibinfo{person}{Marc Swerts}.} \bibinfo{year}{1998}\natexlab{}.
\newblock \showarticletitle{Filled Pauses as Markers of Discourse Structure}.
\newblock \bibinfo{journal}{\emph{Journal of pragmatics}} \bibinfo{volume}{30}, \bibinfo{number}{4} (\bibinfo{year}{1998}), \bibinfo{pages}{485--496}.
\newblock
\urldef\tempurl%
\url{https://doi.org/10.1016/S0378-2166(98)00014-9}
\showDOI{\tempurl}


\bibitem[Taboada(2006)]%
        {taboada2006spontaneous}
\bibfield{author}{\bibinfo{person}{Maite Taboada}.} \bibinfo{year}{2006}\natexlab{}.
\newblock \showarticletitle{Spontaneous and Non-spontaneous Turn-taking}.
\newblock \bibinfo{journal}{\emph{Pragmatics. Quarterly Publication of the International Pragmatics Association (IPrA)}} \bibinfo{volume}{16}, \bibinfo{number}{2-3} (\bibinfo{year}{2006}), \bibinfo{pages}{329--360}.
\newblock
\urldef\tempurl%
\url{https://doi.org/10.1075/prag.16.2-3.04tab}
\showDOI{\tempurl}


\bibitem[Templeton et~al\mbox{.}(2022)]%
        {templeton_fast_2022}
\bibfield{author}{\bibinfo{person}{Emma~M. Templeton}, \bibinfo{person}{Luke~J. Chang}, \bibinfo{person}{Elizabeth~A. Reynolds}, \bibinfo{person}{Marie~D. Cone~LeBeaumont}, {and} \bibinfo{person}{Thalia Wheatley}.} \bibinfo{year}{2022}\natexlab{}.
\newblock \showarticletitle{Fast Response Times Signal Social Connection in Conversation}.
\newblock \bibinfo{journal}{\emph{Proceedings of the National Academy of Sciences of the United States of America}} \bibinfo{volume}{119}, \bibinfo{number}{4} (\bibinfo{date}{Jan.} \bibinfo{year}{2022}), \bibinfo{pages}{e2116915119}.
\newblock
\showISSN{0027-8424}
\urldef\tempurl%
\url{https://doi.org/10.1073/pnas.2116915119}
\showDOI{\tempurl}


\bibitem[Topsakal and Topsakal(2022)]%
        {topsakal_framework_2022}
\bibfield{author}{\bibinfo{person}{Oguzhan Topsakal} {and} \bibinfo{person}{Elif Topsakal}.} \bibinfo{year}{2022}\natexlab{}.
\newblock \showarticletitle{Framework for {A} {Foreign} {Language} {Teaching} {Software} for {Children} {Utilizing} {AR}, {Voicebots} and {ChatGPT} ({Large} {Language} {Models})}.
\newblock \bibinfo{journal}{\emph{The Journal of Cognitive Systems}} \bibinfo{volume}{7}, \bibinfo{number}{2} (\bibinfo{date}{Dec} \bibinfo{year}{2022}), \bibinfo{pages}{33--38}.
\newblock
\showISSN{2548-0650}
\urldef\tempurl%
\url{https://doi.org/10.52876/jcs.1227392}
\showDOI{\tempurl}


\bibitem[Tsai et~al\mbox{.}(2019)]%
        {tsai2019faster}
\bibfield{author}{\bibinfo{person}{Vivian Tsai}, \bibinfo{person}{Timo Baumann}, \bibinfo{person}{Florian Pecune}, {and} \bibinfo{person}{Justine Cassell}.} \bibinfo{year}{2019}\natexlab{}.
\newblock \showarticletitle{Faster {Responses} {Are} {Better} {Responses}: {Introducing} {Incrementality} into {Sociable} {Virtual} {Personal} {Assistants}}. In \bibinfo{booktitle}{\emph{9th {International} {Workshop} on {Spoken} {Dialogue} {System} {Technology}}}, \bibfield{editor}{\bibinfo{person}{Luis~Fernando D'Haro}, \bibinfo{person}{Rafael~E. Banchs}, {and} \bibinfo{person}{Haizhou Li}} (Eds.). \bibinfo{publisher}{Springer}, \bibinfo{address}{Singapore}, \bibinfo{pages}{111--118}.
\newblock
\showISBNx{9789811394430}
\urldef\tempurl%
\url{https://doi.org/10.1007/978-981-13-9443-0_10}
\showDOI{\tempurl}


\bibitem[Vayani et~al\mbox{.}(2024)]%
        {vayani2024languagesmatterevaluatinglmms}
\bibfield{author}{\bibinfo{person}{Ashmal Vayani}, \bibinfo{person}{Dinura Dissanayake}, \bibinfo{person}{Hasindri Watawana}, \bibinfo{person}{Noor Ahsan}, \bibinfo{person}{Nevasini Sasikumar}, \bibinfo{person}{Omkar Thawakar}, \bibinfo{person}{Henok~Biadglign Ademtew}, \bibinfo{person}{Yahya Hmaiti}, \bibinfo{person}{Amandeep Kumar}, {and} \bibinfo{person}{Kartik~Kuckreja et al.}} \bibinfo{year}{2024}\natexlab{}.
\newblock \bibinfo{title}{All Languages Matter: Evaluating LMMs on Culturally Diverse 100 Languages}.
\newblock
\newblock
\showeprint[arxiv]{2411.16508}~[cs.CV]


\bibitem[Villar et~al\mbox{.}(2013)]%
        {villar2013meta}
\bibfield{author}{\bibinfo{person}{Ana Villar}, \bibinfo{person}{Mario Callegaro}, {and} \bibinfo{person}{Yongwei Yang}.} \bibinfo{year}{2013}\natexlab{}.
\newblock \showarticletitle{Where am I? A Meta-analysis of Experiments on the Effects of Progress Indicators for Web Surveys}.
\newblock \bibinfo{journal}{\emph{Social Science Computer Review}} \bibinfo{volume}{31}, \bibinfo{number}{6} (\bibinfo{year}{2013}), \bibinfo{pages}{744--762}.
\newblock
\urldef\tempurl%
\url{https://doi.org/10.1177/0894439313497468}
\showDOI{\tempurl}


\bibitem[Wan et~al\mbox{.}(2024)]%
        {wan_building_2024}
\bibfield{author}{\bibinfo{person}{Hongyu Wan}, \bibinfo{person}{Jinda Zhang}, \bibinfo{person}{Abdulaziz~Arif Suria}, \bibinfo{person}{Bingsheng Yao}, \bibinfo{person}{Dakuo Wang}, \bibinfo{person}{Yvonne Coady}, {and} \bibinfo{person}{Mirjana Prpa}.} \bibinfo{year}{2024}\natexlab{}.
\newblock \showarticletitle{Building {LLM}-based {AI} {Agents} in {Social} {Virtual} {Reality}}. In \bibinfo{booktitle}{\emph{Extended {Abstracts} of the 2024 {CHI} {Conference} on {Human} {Factors} in {Computing} {Systems}}} \emph{(\bibinfo{series}{{CHI} {EA} '24})}. \bibinfo{publisher}{Association for Computing Machinery}, \bibinfo{address}{New York, NY, USA}, \bibinfo{pages}{1--7}.
\newblock
\showISBNx{9798400703317}
\urldef\tempurl%
\url{https://doi.org/10.1145/3613905.3651026}
\showDOI{\tempurl}


\bibitem[Wang et~al\mbox{.}(2024)]%
        {wang_virtuwander_2024}
\bibfield{author}{\bibinfo{person}{Zhan Wang}, \bibinfo{person}{Lin-Ping Yuan}, \bibinfo{person}{Liangwei Wang}, \bibinfo{person}{Bingchuan Jiang}, {and} \bibinfo{person}{Wei Zeng}.} \bibinfo{year}{2024}\natexlab{}.
\newblock \showarticletitle{{VirtuWander}: {Enhancing} {Multi}-modal {Interaction} for {Virtual} {Tour} {Guidance} through {Large} {Language} {Models}}. In \bibinfo{booktitle}{\emph{Proceedings of the {CHI} {Conference} on {Human} {Factors} in {Computing} {Systems}}} \emph{(\bibinfo{series}{{CHI} '24})}. \bibinfo{publisher}{Association for Computing Machinery}, \bibinfo{address}{New York, NY, USA}, \bibinfo{pages}{1--20}.
\newblock
\showISBNx{9798400703300}
\urldef\tempurl%
\url{https://doi.org/10.1145/3613904.3642235}
\showDOI{\tempurl}


\bibitem[Weizenbaum(1966)]%
        {weiz1966eliza}
\bibfield{author}{\bibinfo{person}{Joseph Weizenbaum}.} \bibinfo{year}{1966}\natexlab{}.
\newblock \showarticletitle{ELIZA—a computer program for the study of natural language communication between man and machine}.
\newblock \bibinfo{journal}{\emph{Commun. ACM}} \bibinfo{volume}{9}, \bibinfo{number}{1} (\bibinfo{date}{Jan.} \bibinfo{year}{1966}), \bibinfo{pages}{36–45}.
\newblock
\showISSN{0001-0782}
\urldef\tempurl%
\url{https://doi.org/10.1145/365153.365168}
\showDOI{\tempurl}


\bibitem[Westerman et~al\mbox{.}(2019)]%
        {westerman_i_2019}
\bibfield{author}{\bibinfo{person}{David Westerman}, \bibinfo{person}{Aaron~C. Cross}, {and} \bibinfo{person}{Peter~G. Lindmark}.} \bibinfo{year}{2019}\natexlab{}.
\newblock \showarticletitle{I {Believe} in a {Thing} {Called} {Bot}: {Perceptions} of the {Humanness} of “{Chatbots}”}.
\newblock \bibinfo{journal}{\emph{Communication Studies}} \bibinfo{volume}{70}, \bibinfo{number}{3} (\bibinfo{date}{May} \bibinfo{year}{2019}), \bibinfo{pages}{295--312}.
\newblock
\showISSN{1051-0974}
\urldef\tempurl%
\url{https://doi.org/10.1080/10510974.2018.1557233}
\showDOI{\tempurl}


\bibitem[Wigdor et~al\mbox{.}(2016)]%
        {wigdor2016improve}
\bibfield{author}{\bibinfo{person}{Noel Wigdor}, \bibinfo{person}{Joachim de Greeff}, \bibinfo{person}{Rosemarijn Looije}, {and} \bibinfo{person}{Mark~A Neerincx}.} \bibinfo{year}{2016}\natexlab{}.
\newblock \showarticletitle{How to Improve Human-robot Interaction with Conversational Fillers}. In \bibinfo{booktitle}{\emph{2016 25th IEEE international symposium on robot and human interactive communication (RO-MAN)}}. IEEE, \bibinfo{publisher}{IEEE}, \bibinfo{address}{New York, NY, USA}, \bibinfo{pages}{219--224}.
\newblock
\urldef\tempurl%
\url{https://doi.org/10.1109/ROMAN.2016.7745134}
\showDOI{\tempurl}


\bibitem[Wobbrock et~al\mbox{.}(2011)]%
        {wobbrock2011aligned}
\bibfield{author}{\bibinfo{person}{Jacob~O. Wobbrock}, \bibinfo{person}{Leah Findlater}, \bibinfo{person}{Darren Gergle}, {and} \bibinfo{person}{James~J. Higgins}.} \bibinfo{year}{2011}\natexlab{}.
\newblock \showarticletitle{The aligned rank transform for nonparametric factorial analyses using only anova procedures}. In \bibinfo{booktitle}{\emph{Proceedings of the SIGCHI Conference on Human Factors in Computing Systems}} (Vancouver, BC, Canada) \emph{(\bibinfo{series}{CHI '11})}. \bibinfo{publisher}{Association for Computing Machinery}, \bibinfo{address}{New York, NY, USA}, \bibinfo{pages}{143–146}.
\newblock
\showISBNx{9781450302289}
\urldef\tempurl%
\url{https://doi.org/10.1145/1978942.1978963}
\showDOI{\tempurl}


\bibitem[Wu et~al\mbox{.}(2022)]%
        {wu2022time}
\bibfield{author}{\bibinfo{person}{Charley Wu}, \bibinfo{person}{Eric Schulz}, \bibinfo{person}{Timothy Pleskac}, {and} \bibinfo{person}{Maarten Speekenbrink}.} \bibinfo{year}{2022}\natexlab{}.
\newblock \showarticletitle{Time pressure changes how people explore and respond to uncertainty}.
\newblock \bibinfo{journal}{\emph{Scientific Reports}}  \bibinfo{volume}{12} (\bibinfo{date}{March} \bibinfo{year}{2022}), \bibinfo{pages}{4122}.
\newblock
\urldef\tempurl%
\url{https://doi.org/10.1038/s41598-022-07901-1}
\showDOI{\tempurl}


\bibitem[Yamazaki et~al\mbox{.}(2023)]%
        {yamazaki_open-domain_2023}
\bibfield{author}{\bibinfo{person}{Takato Yamazaki}, \bibinfo{person}{Tomoya Mizumoto}, \bibinfo{person}{Katsumasa Yoshikawa}, \bibinfo{person}{Masaya Ohagi}, \bibinfo{person}{Toshiki Kawamoto}, {and} \bibinfo{person}{Toshinori Sato}.} \bibinfo{year}{2023}\natexlab{}.
\newblock \showarticletitle{An {Open}-{Domain} {Avatar} {Chatbot} by {Exploiting} a {Large} {Language} {Model}}. In \bibinfo{booktitle}{\emph{Proceedings of the 24th {Annual} {Meeting} of the {Special} {Interest} {Group} on {Discourse} and {Dialogue}}}, \bibfield{editor}{\bibinfo{person}{Svetlana Stoyanchev}, \bibinfo{person}{Shafiq Joty}, \bibinfo{person}{David Schlangen}, \bibinfo{person}{Ondrej Dusek}, \bibinfo{person}{Casey Kennington}, {and} \bibinfo{person}{Malihe Alikhani}} (Eds.). \bibinfo{publisher}{Association for Computational Linguistics}, \bibinfo{address}{Prague, Czechia}, \bibinfo{pages}{428--432}.
\newblock
\urldef\tempurl%
\url{https://doi.org/10.18653/v1/2023.sigdial-1.40}
\showDOI{\tempurl}


\bibitem[Yang and Dorneich(2015)]%
        {yang2015effect}
\bibfield{author}{\bibinfo{person}{Euijung Yang} {and} \bibinfo{person}{Michael~C. Dorneich}.} \bibinfo{year}{2015}\natexlab{}.
\newblock \showarticletitle{The {Effect} of {Time} {Delay} on {Emotion}, {Arousal}, and {Satisfaction} in {Human}-{Robot} {Interaction}}.
\newblock \bibinfo{journal}{\emph{Proceedings of the Human Factors and Ergonomics Society Annual Meeting}} \bibinfo{volume}{59}, \bibinfo{number}{1} (\bibinfo{date}{Sept.} \bibinfo{year}{2015}), \bibinfo{pages}{443--447}.
\newblock
\showISSN{1071-1813}
\urldef\tempurl%
\url{https://doi.org/10.1177/1541931215591094}
\showDOI{\tempurl}
\newblock
\shownote{Publisher: SAGE Publications Inc}.


\bibitem[Yang et~al\mbox{.}(2024)]%
        {yang_effects_2024}
\bibfield{author}{\bibinfo{person}{Fu-Chia Yang}, \bibinfo{person}{Kevin Duque}, {and} \bibinfo{person}{Christos Mousas}.} \bibinfo{year}{2024}\natexlab{}.
\newblock \showarticletitle{The {Effects} of {Depth} of {Knowledge} of a {Virtual} {Agent}}.
\newblock \bibinfo{journal}{\emph{IEEE Transactions on Visualization and Computer Graphics}} \bibinfo{volume}{30}, \bibinfo{number}{11} (\bibinfo{date}{Nov.} \bibinfo{year}{2024}), \bibinfo{pages}{7140--7151}.
\newblock
\showISSN{1077-2626, 1941-0506, 2160-9306}
\urldef\tempurl%
\url{https://doi.org/10.1109/TVCG.2024.3456148}
\showDOI{\tempurl}


\bibitem[Yu et~al\mbox{.}(2020)]%
        {yu_engaging_2020}
\bibfield{author}{\bibinfo{person}{Difeng Yu}, \bibinfo{person}{Qiushi Zhou}, \bibinfo{person}{Benjamin Tag}, \bibinfo{person}{Tilman Dingler}, \bibinfo{person}{Eduardo Velloso}, {and} \bibinfo{person}{Jorge Goncalves}.} \bibinfo{year}{2020}\natexlab{}.
\newblock \showarticletitle{Engaging {Participants} during {Selection} {Studies} in {Virtual} {Reality}}. In \bibinfo{booktitle}{\emph{2020 {IEEE} {Conference} on {Virtual} {Reality} and {3D} {User} {Interfaces} ({VR})}}. \bibinfo{publisher}{IEEE}, \bibinfo{address}{Atlanta, GA, USA}, \bibinfo{pages}{500--509}.
\newblock
\urldef\tempurl%
\url{https://doi.org/10.1109/VR46266.2020.00071}
\showDOI{\tempurl}
\newblock
\shownote{ISSN: 2642-5254}.


\bibitem[Zhu et~al\mbox{.}(2023)]%
        {zhu_free_form_2023}
\bibfield{author}{\bibinfo{person}{Jiarui Zhu}, \bibinfo{person}{Radha Kumaran}, \bibinfo{person}{Chengyuan Xu}, {and} \bibinfo{person}{Tobias Höllerer}.} \bibinfo{year}{2023}\natexlab{}.
\newblock \showarticletitle{Free-form {Conversation} with {Human} and {Symbolic} {Avatars} in {Mixed} {Reality}}. In \bibinfo{booktitle}{\emph{2023 {IEEE} {International} {Symposium} on {Mixed} and {Augmented} {Reality} ({ISMAR})}}. \bibinfo{publisher}{IEEE}, \bibinfo{address}{Sydney, Australia}, \bibinfo{pages}{751--760}.
\newblock
\urldef\tempurl%
\url{https://doi.org/10.1109/ISMAR59233.2023.00090}
\showDOI{\tempurl}
\newblock
\shownote{ISSN: 2473-0726}.


\end{thebibliography}


\end{document}